\documentclass[aps,prb,reprint,superscriptaddress,nobalancelastpage,twocolumn,showpacs]{revtex4-1}
\usepackage{color}
\usepackage{comment}
\usepackage{graphicx}
\usepackage{amsmath}
\usepackage{latexsym}
\usepackage{epstopdf}
\usepackage{amsmath,amssymb}
\usepackage{amssymb,amsmath}
\usepackage{multirow}
\usepackage{mathtools}
\usepackage[dvipsnames]{xcolor}
\newcommand{\beq}{\begin{equation}}
\newcommand{\eeq}{\end{equation}}

\usepackage{tikz}
\usepackage{xcolor}
\usetikzlibrary{patterns}

\begin{document}

\title{Light-cone velocities  after a global quench in a non-interacting model }

\author{K. Najafi  }
\affiliation{Department of Physics, Georgetown University,
37th and O Sts.  NW, Washington, DC 20057, USA}
\author{ M.~A.~Rajabpour}

\affiliation{  Instituto de F\'isica, Universidade Federal Fluminense, Av. Gal. Milton Tavares de Souza s/n, Gragoat\'a, 24210-346, Niter\'oi, RJ, Brazil}
\date{\today{}}

\author{ J. Viti}

\affiliation{ ECT \& Instituto Internacional de F\'isica, UFRN,
Campos Universit\'ario, Lagoa Nova 59078-970 Natal, Brazil}
\date{\today{}}

\begin{abstract}

We study  the  light-cone velocity for global quenches in the non-interacting XY chain  starting from a class of initial states that are eigenstates
of the local $z$-component of the spin. We point out
how translation invariance of the initial state can affect the maximal speed at
which correlations spread. 
As a consequence the light-cone velocity can be state-dependent also for non-interacting systems:
a new effect of which we provide clear numerical evidence and analytic predictions.  Analogous considerations, based on numerical results, are drawn for the evolution of the entanglement entropy. 
\end{abstract}
\maketitle

\section{Introduction}
In a seminal contribution Lieb and Robinson~\cite{LR}
proved that in a non-relativistic quantum  spin system,
the operator norm of the commutator between two local observables
$\hat{A}(\boldsymbol{x},t)$  and $\hat{B}(\boldsymbol{y},0)$  is exponentially small
as long as $|\boldsymbol{x}-\boldsymbol{y}|\geq v_{LR} t$, for some $v_{LR}>0$~\cite{BHV, EO, NS}. The bound reads
\begin{equation} \label{LR bound}
||[\hat{A}(\boldsymbol{x},t),\hat{B}(\boldsymbol{y},0)]||\leq c||\hat{A}||||\hat{B}|| e^{-(|\boldsymbol{x}-\boldsymbol{y}|-v_{LR} t)/\xi},
\end{equation}
where $||\hat{O}||$ is the aforementioned operator norm of the observable $\hat{O}$; $c$ and $\xi$ are constants.
The Lieb-Robinson theorem holds in any dimension
and for a translation invariant Hamiltonian with interactions decaying sufficiently fast (exponentially or superexponentially) with the distance.
The parameter $v_{LR}$ is called  Lieb-Robinson velocity and  depends only on the Hamiltonian~\cite{LR, BHV, EO, NS}
driving the time evolution; in particular it is state-independent\cite{LRvelocity}.
The existence of a finite $v_{LR}$
implies that information cannot propagate arbitrary fast~\cite{LR}. The Lieb-Robinson result
represents nowadays a powerful tool to prove rigorous bounds for  correlation
and entanglement growth~\cite{E2010}. It is relevant for estimating the computational complexity, or simulability, of a system~\cite{Has2009, E2010} and
provides a link between the presence of a gap and short length
correlations~\cite{NS, HK, NS2}.

Moreover, formidable experimental~\cite{K, Sch2007, BDZ2008, Isingex, SGibbs}
and theoretical (for an overview~\cite{P2011, GE2015,Rigol2016, intro}) progress in many-body quantum
dynamics, have further underlined
its striking physical implications. Among them,  the emergence of light-cone effects in 
correlation functions of time-evolved local observables for systems that are not manifestly
Lorentz invariant, like quantum  spin chains~\cite{LK2008, MWNM2009, CEF2011, CEF2011a, CEF2011b, B2016, DSC2016,CDCTS2017,Porta2016,Porta2018}. For instance, the Lieb-Robinson theorem  has been invoked
to explain the ballistic spreading of correlation functions in paradigmatic
condensed matter models such as the XXZ spin chain~\cite{Essler2014} or the
Hubbard model~\cite{CF2014} after a global
quench; see also~\cite{EF2016} for a pedagogical survey. Similar considerations apply  to the linear growth
of the entanglement entropies~\cite{CC2005, DMCF2006, FC2008, KH, SD2011, GR14, NRV, D2017, CHN}.
 Quantitative predictions on the spreading of correlations
and the entanglement entropy are also of clear experimental interest; see~\cite{CBPE2012, LS2013, exp3, zoll} for experimental verifications of light-cone effects
in many-body systems. 
A complementary physical interpretation of these emergent phenomena is  based
on the idea~\cite{CC2006, CC2007} that in a quench problem the initial state acts as a source of pairs
of entangled quasi-particles. The quasi-particles have opposite momenta and
move ballistically; see~\cite{AC2017, AC2017b, PKZ} for applications to integrable models.
Recently it has also  been pointed out that
light-cone effects can be recovered from field theoretical
arguments without relaying on particular properties of the post-quench quasi-particle
dynamics~\cite{A2018}.

It is  important to stress that the Lieb-Robinson theorem proves the
existence of a maximal
velocity for correlations to develop.
However the observed propagation velocity  is actually  state-dependent and non-trivially predictable. In other words,
it is not directly related to $v_{LR}$ that rather furnishes an upper bound.  For one-dimensional 
spin chains, a state-dependent light-cone velocity was noticed first in~\cite{Essler2014}. In a context of integrability, 
it was pointed out how the dispersions of the quasi-particles in the stationary state~\cite{CE2013} (the so-called
Generalized Gibbs Ensemble~\cite{R1, R2}) was initial state dependent.
A prediction for their velocities  was then proposed and numerically tested. This idea has lead to many
subsequent crucial developments in the field~\cite{hydro, hydro2, hydro3, AC2017, AC2017b}. 
In the simpler setting of a non-interacting model, where quasi-particles are characterized by a dispersion relation $\varepsilon_k$ 
there is a  maximum group velocity $v_g$ allowed by the dispersion. To our best knowledge, in all the examples considered
in the literature so-far (for example ~\cite{CEF2011,CEF2011a, CEF2011b,Bucciantini2014a,Bucciantini2014b}) 
correlations were found to spread with a light-cone velocity given by $v_{g}$ independently of the initial state.
In this paper we  point out for the first time  that 
even if  the dispersion of the quasi-particles after the quench is not affected by the initial state, 
its symmetries 
and in particular translation invariance can
introduce additional selection rules on the momenta of the quasi-particles.  As a consequence the light-cone velocity can be state-dependent and smaller than $v_g$ also in a non-interacting model

The paper is organized as follows. In Sec.~\ref{sec2} and Sec.~\ref{sec3} we introduce the XY spin chain and discuss a special class of initial states
for which  time-evolution of local observables can be calculated easily. In Sec.~\ref{sec4} and~\ref{sec5} we analyze the propagation velocity both for physical
two-point correlation functions and the entanglement entropy. In Sec.~\ref{app}
we show how fermionic correlation functions
at the edge of the light-cone can be approximated by combinations of Airy functions.
Similar statements were also made in~\cite{CAiry, B2016} for  related many-body quench problems. 
After the conclusions in Sec.~\ref{conc}, one appendix completes the paper. 


\section{XY chain: notations and initial states}
\label{sec2}

In this section, we  briefly remind the XY chain, its fermionic representation and we then introduce the initial states discussed in the rest of the paper.
The Hamiltonian of the XY-chain~\cite{LSM} is 
\begin{align}
\nonumber\textbf{H}_{XY}=&-J\sum_{l=1}^{L}\Big{[}(\frac{1+\gamma}{2})\sigma_{l}^{x}\sigma_{l+1}^{x}+(\frac{1-\gamma}{2})\sigma_{l}^{y}\sigma_{l+1}^{y}\Big{]}\\
&\label{HXY1} -Jh\sum_{l=1}^{L}\sigma_{l}^{z},
\end{align} 
where the $\sigma_{l}^{\alpha}$ ($\alpha=x,y,z$) are Pauli matrices and $J,\gamma,h$ are real parameters; in particular $J>0$ and conventionally $h$ is called magnetic field. The
XY model reduces to the Ising spin chain for $\gamma=1$ and it is the XX chain when $\gamma=0$. We will also choose periodic boundary conditions for the spins,
i.e. $\sigma_{l}^{\alpha}=\sigma_{l+L}^{\alpha}$. 
Introducing canonical spinless fermions through the Jordan-Wigner transformation,  $c_l^{\dagger}=\prod_{n<l}\sigma_n^z\sigma_l^{+}$, \eqref{HXY1} becomes 
\begin{equation}\label{XY-ff}
\textbf{H}_{XY}=J\sum_{l=1}^{L} (c_l^{\dagger}c_{l+1}+\gamma c_l^{\dagger}c_{l+1}^{\dagger}+h.c.)
-Jh\sum_{l=1}^{L}(2c_l^{\dagger}c_l-1)
\end{equation}
where $c_{L+1}^{\dagger}=-\mathcal{N}c_{1}^{\dagger}$. Here $\mathcal{N}=\pm1$ is the eigenvalue of operator $\prod_{l=1}^{L}\sigma_l^z$ that is conserved fermion parity.
The above Hamiltonian can be written as %
\begin{eqnarray}\label{H1}\
\textbf{H}_{XY}=\textbf{c}^{\dagger}\textbf{A}\textbf{c}+\frac{1}{2}\textbf{c}^{\dagger}\textbf{B}\textbf{c}^{\dagger}+\frac{1}{2}\textbf{c}\textbf{B}^{T}\textbf{c}
-\frac{1}{2}{\rm Tr}{\textbf{A}},
\end{eqnarray} 
with appropriate real matrices $\textbf{A}$ and $\textbf{B}$ that are symmetric and antisymmetric, respectively.    
In the sector with an even number of fermions ($\mathcal{N}=1$), the so-called Neveau-Schwartz sector, the Hamiltonian can be
diagonalized in Fourier space by a unitary Bogoliubov transformation. In particular, there are no subtleties related
to the appearance of a zero-mode. Writing (\ref{XY-ff}) in the Fourier space and then Bogolubov transformation leads to
\begin{eqnarray}\label{H diagonal}\
\textbf{H}_{XY}=\sum_{k=1}^{L}\varepsilon_k(b_k^{\dagger}b_k-\frac{1}{2}).
\end{eqnarray}
The canonical Bogoliubov fermions $b$'s have the following dispersion relation and group velocities
\begin{align}\label{dispersion}
&\varepsilon_k=2J\sqrt{(\cos\phi_k-h)^2+\gamma^2\sin^2\phi_k},\\
\label{group velocity}
&v_k=2J\sin\phi_k\frac{\gamma^2\cos\phi_k-\cos\phi_k+h}{\sqrt{(\cos\phi_k-h)^2+\gamma^2\sin^2\phi_k}},
\end{align} 
where $\phi_k=\frac{2\pi}{L}(k-\frac{1}{2})$ and $k=1,...,L$ . We will assume $L$ even from now on. The diagonalization procedure will be also briefly revisited in Sec.~\ref{sec3}.

In this paper, we are interested to study the time evolution  of such a system from initial states  that are eigenstates of the local $\sigma_{j}^z$ operators.
For example, the initial state $|\psi_0\rangle$ can be a state with all spins pointing up (no fermions) or down (one fermion per lattice site); other possibilities can be the N\'eel state $|\downarrow\uparrow\downarrow\uparrow...\downarrow\uparrow\rangle$ and alike.
Moreover, all the states that we study have a  periodic pattern in real space with a fixed number of spin up. We label our crystalline
initial states $|\psi_0\rangle$ as $(r,s)$, where $r$ is the spin up (fermion) density  and $s$ is the number of spin up
in the unit cell of the crystal. For example, the N\'eel state is labeled by $(\frac{1}{2},1)$ and the state $|\downarrow\downarrow\uparrow\uparrow\downarrow\downarrow...\downarrow\downarrow\uparrow\uparrow\rangle$ will be
$(\frac{1}{2},2)$. It is also convenient to define $p\equiv\frac{s}{r}$ that for simplicity we restrict to be a positive integer
(i.e. $s$ is a multiple of $r$).  Although the class of initial state considered is not comprehensive,
it turns out to be enough for the upcoming discussion.

\section{Evolution of the correlation functions}
\label{sec3}
 Light-cone effects can be studied, monitoring the connected correlation function  of the $z$-component of the spin $S^z=\sigma^z/2$,
\begin{equation}
\label{corrdef}
\Delta_{ln}(t)=\langle S^z_l(t)S^z_n(t)\rangle-\langle S^z_l(t)\rangle \langle S^z_n(t)\rangle.
\end{equation} 
For our initial states, \eqref{corrdef} is zero at time $t=0$; however, according to~\cite{LR}, after a certain time
which depends on $|l-n|$, it starts to change significantly.

For instance~\cite{Essler2014}, such a time can be chosen as the first inflection point $\tau$. Varying $|l-n|$ in Eq.~\eqref{corrdef}, one can numerically evaluate  $\tau$ and provide a prediction  for the
speed $v_{\text{max}}$  at which information spreads in the system 
determining the ratio $\frac{|l-n|}{2\tau}$.
We will call $v_{\text{max}}$ the light-cone velocity.
According to a quasi-particle picture of the quench dynamics~\cite{CC2006,CC2007,CEF2011,CEF2011a, CEF2011b,Bucciantini2014a,Bucciantini2014b, AC2017, AC2017b},
for a translation invariant Hamiltonian the initial state $|\psi_0\rangle$ acts as a source of pairs of entangled quasi-particles with
opposite momenta. In absence of interactions, the quasi-particles
move ballistically with a group velocity fixed by their dispersion relation. Within this  framework,  $\tau$ is state-independent and $v_{\max}=v_g$ where
 $v_g>0$ is the maximum over the $k$'s of Eq.~\eqref{group velocity}
We will actually show that $v_g$  is rather an upper bound for the  observed $v_{\text{max}}$, which  can indeed be dependent on  the symmetries of the initial state. Finally note 
that in absence of interactions, the light-cone velocity is independent on finite size effects as long as  $L\gg |l-n|$.

To study the time-evolution of the  correlation function \eqref{corrdef}, first, we need to analyze the propagators
\begin{align}\label{flndef}
F_{ln}(t)&=\langle\psi_0 | c_{l}^{\dagger}(t)c_{n}^{\dagger}(t)|\psi_0 \rangle,\\
\label{clndef}
C_{ln}(t)&=\langle \psi_0 |c_{l}(t)c_{n}^{\dagger}(t)|\psi_0\rangle,
\end{align}
where $|\psi_0 \rangle$ is a state of the type introduced in Sec.~\ref{sec2}. If there is no ambiguity,  we will drop it from the expectation values.
From Eqs. \eqref{flndef}-\eqref{clndef} and the Wick theorem, which applies to our states~\cite{foot}, it follows
\begin{equation}\label{ correlations}
\Delta_{ln}(t)=|F_{ln}(t)|^2-|C_{ln}(t)|^2.
\end{equation}

Note that as $L\times L$ matrices whose matrix elements are given in Eqs. \eqref{flndef}-\eqref{clndef}, $\boldsymbol{F}$ and $\boldsymbol{C}$ 
satisfy  $\textbf{F}^{T}=-\textbf{F}$ and $\textbf{C}^{\dagger}=\textbf{C}$.

It is straightforward to show that for a fermionic model with  a quadratic Hamiltonian as in Eq. (\ref{H1}) one has,
\begin{equation}\label{clcn1}\
\begin{pmatrix}
\textbf{c}(t) \\
\textbf{c}^{\dagger}(t)\\
  \end{pmatrix} =\underbrace{e^{-it\begin{pmatrix}
\textbf{A} & \textbf{B}\\
\textbf{-B} & \textbf{-A}\\
  \end{pmatrix}}}_{\boldsymbol{T}(t)}
  \begin{pmatrix}
  \textbf{c}(0) \\
\textbf{c}^{\dagger}(0)\\
 \end{pmatrix}
\end{equation} 
where $\textbf{c}=\{c_1,c_2,...,c_L\}$ and $\textbf{c}^{\dagger}=\{c^{\dagger}_1,c^{\dagger}_2,...,c^{\dagger}_L\}$ are vectors of length $L$; i.e. $\boldsymbol{T}$ is a $2L\times 2L$ matrix.
\begin{table}[t!]
\centering
{\begin{tabular}{|l|c|}
  \hline 
\textbf{Generic Quadratic Hamiltonian}\\
  \hline
$\textbf{T}_{11}$, $\textbf{T}_{12}$, $\textbf{T}_{21}$ and $\textbf{T}_{22}$ are complex matrices.\\
   \hline
$\textbf{T}_{21}=(\textbf{T}_{12})^*$\\
   \hline
$\textbf{T}_{22}=(\textbf{T}_{11})^*$\\
\hline
$(\textbf{T}_{11})^{T}=\textbf{T}_{11}$\\
\hline
$(\textbf{T}_{22})^{T}=\textbf{T}_{22}$\\
\hline
$\textbf{T}_{22}\textbf{T}_{12}+\textbf{T}_{21}\textbf{T}_{22}=0$\\
\hline
\end{tabular}}
\label{table1}
\caption{Properties of the four $L\times L$ blocks of the matrix $\textbf{T}$ for a quadratic Hamiltonian \eqref{H1}. The notation is obvious and time dependence is omitted here.} 
\end{table} 
Exploiting the properties of the four $L\times L$ blocks of the matrix $\textbf{T}$  collected in Tab.~\ref{table1}, Eq.~\eqref{clcn1} can be written as,
\begin{eqnarray}\label{clcn2}\
\begin{pmatrix}
\textbf{c}(t) \\
\textbf{c}^{\dagger}(t)\\
  \end{pmatrix} =\begin{pmatrix}
\textbf{T}_{22}^{*} (t) & \textbf{T}_{12}(t)\\
\textbf{T}_{12}^{*}(t) & \textbf{T}_{22}(t)\\
  \end{pmatrix} \begin{pmatrix}
  \textbf{c}(0) \\
\textbf{c}^{\dagger}(0)\\
 \end{pmatrix}.
\end{eqnarray} 

Finally, after some easy algebra, explicit expression for the time-evolved matrices $\boldsymbol{F}$ and $\boldsymbol{C}$ can be computed and read respectively (time dependence is dropped from the $\boldsymbol{T}$'s)
\begin{align}\label{fl2}
\textbf{F}(t)=&\textbf{T}_{12}^{*}\textbf{F}^{\dagger}(0)\textbf{T}_{12}^{\dagger}+\textbf{T}_{12}^{*}\textbf{C}(0)\textbf{T}_{22}+\textbf{T}_{22}\textbf{T}_{12}^{\dagger} \nonumber \\  &-\textbf{T}_{22}\textbf{C}^{T}(0)\textbf{T}_{12}^{\dagger}+\textbf{T}_{22}\textbf{F}(0)\textbf{T}_{22},
\end{align}
\begin{align}\label{cl2}
\textbf{C}(t)=&\textbf{T}_{22}^{*}\textbf{F}^{\dagger}(0)\textbf{T}_{12}^{\dagger}+\textbf{T}_{22}^{*}\textbf{C}(0)\textbf{T}_{22}+\textbf{T}_{12}\textbf{T}_{12}^{\dagger}\nonumber \\  &-\textbf{T}_{12}\textbf{C}^{T}(0)\textbf{T}_{12}^{\dagger}+\textbf{T}_{12}\textbf{F}(0)\textbf{T}_{22}.
\end{align}
Eqs. \eqref{fl2}-\eqref{cl2} are valid in principle for any free fermionic systems with Hamiltonian \eqref{H1}; however,  they 
have much simpler forms in the XY chain for our particular choice of initial states as we now discuss.

In the periodic XY chain, it turns out  $[\textbf{A},\textbf{B}]=0$; these matrices are indeed trivially diagonalized by the unitary transformation with elements $U_{lk}=\frac{1}{\sqrt{L}}e^{-il\phi_k}$ and $\phi_k$ given, for $\mathcal N=1$, below Eq.~\eqref{group velocity}. Consequently, the four blocks  of $\textbf{T}$  are mutually commuting  and this leads to further simplifications. In particular, it is easy to verify that
\begin{align}\label{T11}
&\textbf{T}_{11}=\cos[t\sqrt{\textbf{A}^2-\textbf{B}^2}]-
\frac{i\textbf{A}}{\sqrt{\textbf{A}^2-\textbf{B}^2}}\sin[t\sqrt{\textbf{A}^2-\textbf{B}^2}],\\
\label{T12}
&\textbf{T}_{12}=\frac{-i \textbf{B}}{\sqrt{\textbf{A}^2-\textbf{B}^2}}\sin[t\sqrt{\textbf{A}^2-\textbf{B}^2}].
\end{align}
The other two blocks can be found observing that $\textbf{T}_{22}=\textbf{T}_{11}^{*}$ and $\textbf{T}_{21}=-\textbf{T}_{12}$.
The eigenvalues of the matrices $\mathbf{A}$ and $\mathbf{B}$ are
\begin{align}\label{eigenvalues1}
&\lambda_k^{A}=2J(-h+\cos\phi_k),\\
\label{eigenvalues2}
&\lambda_k^{B}=-2iJ\gamma \sin\phi_k.
\end{align}
From Eqs.~\eqref{eigenvalues1}-\eqref{eigenvalues2} and comparing with Eq.~\eqref{dispersion}, it follows $\varepsilon_k=\sqrt{(\lambda_k^{A})^2-(\lambda_k^{B})^2}$. Recalling then Eqs.(\ref{T11})-(\ref{T12}) we finally obtain 
\begin{align}
\label{T22b}
&\lambda_{k}^{T_{11}}=[\lambda_{k}^{T_{22}}]^{*}=\cos(t\varepsilon_{k})-i\frac{\lambda_{k}^A}{\epsilon_{k}}\sin(t\varepsilon_{k})\\
\label{T12b}
&\lambda_{k}^{T_{12}}=-\lambda_{k}^{T_{21}}=-i\frac{\lambda_{k}^{B}}{\varepsilon_{k}}\sin(t\varepsilon_{k}).
\end{align}
The time-evolved matrices $\textbf{F}(t)$ and $\textbf{C}(t)$ can now be calculated for the class of representative initial states $|\psi_0\rangle$ introduced in Sec.~\ref{sec2}. A trivial case is  $|\psi_0\rangle=|\downarrow\downarrow\dots \downarrow\downarrow\rangle$, i.e. a state without fermions; here $\mathbf{F}(0)=\mathbf{0}$ and $\textbf{C}(0)=\boldsymbol{1}$. Then using unitarity of the matrix $\boldsymbol{T}$ we obtain
\begin{align}\label{nofermions1}
&[\textbf{F}(t)]_{ln}=
\frac{1}{L}\sum_{k=1}^L\lambda_k^{T_{22}}\lambda_k^{T_{12}}e^{-i\phi_k(l-n)}\\
\label{nofermions2}
&[\textbf{C}(t)]_{ln}=\delta_{ln}-\frac{1}{L}\sum_{k=1}^L(\lambda_k^{T_{12}})^{2}e^{-i\phi_k(l-n)}.
\end{align}
According to Eqs.~\eqref{T22b}-\eqref{nofermions2}, the time evolution of $\Delta_{ln}(t)$ in Eq.~\eqref{corrdef} is described by the ballistic spreading of quasi-particles with dispersion relation $\varepsilon_k$ as in \eqref{dispersion}. As it can be also easily checked numerically, correlations  spread at  the maximum group velocity $v_g$ obtained from (\ref{group velocity}), in agreement with a standard quasi-particle interpretation.

For the  initial states labeled by $(1/p,1)$, the matrix $\textbf{C}(0)$ has elements
\begin{equation}\label{in_corr}
[\boldsymbol{C}(0)]_{ln}=\delta_{ln}\left[1-\frac{1}{p}\sum _{j=0}^{p-1}e^{-2\pi i \frac{lj}{p}}\right],
\end{equation}
whereas $\mathbf{F}(0)=\bf{0}$. From the expressions in Eqs.~\eqref{fl2}-\eqref{cl2} and inserting the unitary matrix $U$ that diagonalize simultaneously all four blocks of $\mathbf{T}$ we derive 

\begin{multline}\label{fln_terms2}
[\textbf{F}(t)]_{ln}=\frac{1}{L}\sum_{k=1}^L\lambda_k^{T_{12}^*}~\lambda_k^{T_{22}}e^{-i\phi_k(l-n)}\\
-\frac{1}{Lp}\sum_{j=0}^{p-1}e^{-\frac{2\pi inj}{p}}
\sum_{k=1}^L\lambda_k^{T_{12}^{*}}~\lambda_{-\frac{Lj}{p}+k}^{T_{22}}e^{-i\phi_k
(l-n)}\\
+\frac{1}{Lp}\sum_{j=0}^{p-1}e^{-\frac{2\pi inj}{p}}\sum_{k=1}^L\lambda_k^{T_{22}}~\lambda_{-\frac{Lj}{p}+k}^{T_{12}}e^{-i\phi_k(l-n)},
\end{multline}
and
\begin{multline}\label{cln_terms2}
[\textbf{C}(t)]_{ln}=\frac{1}{L}\sum_{k=1}^L|\lambda_k^{T_{22}}|^2e^{-i\phi_k(l-n)}
\\
-\frac{1}{Lp}\sum_{j=0}^{p-1}e^{-\frac{2\pi inj}{p}}\sum_{k=1}^L\lambda_k^{T_{22}^{*}}~\lambda_{-\frac{Lj}{p}+k}^{T_{22}}e^{-i\phi_k(l-n)}\\
+\frac{1}{Lp}\sum_{j=0}^{p-1}e^{-\frac{2\pi inj}{p}}\sum_{k=1}^L\lambda_k^{T_{12}}~\lambda_{-\frac{Lj}{p}+k}^{T_{12}}e^{-i\phi_k(l-n)}.
\end{multline}
 It should be noticed that for $p=2$ the terms in the first lines of Eqs.~\eqref{fln_terms2}-\eqref{cln_terms2}  are canceled by the $j=0$ contribution
in the sum.
The latter observation follows from  
\begin{equation}\label{ T22 and T12 equality}
\lambda_k^{T^*_{22}}\lambda_{-k}^{T_{22}}+(\lambda_k^{T_{12}})^{2}=1,
\end{equation}
that is a consequence of the unitarity of the matrix $\boldsymbol{T}$ and it will be useful in our analysis of the light-cone velocities.
In the next  section, we will pass to study Eqs.~\eqref{fln_terms2}-\eqref{cln_terms2} in details. 

A similar study of the Eq.~\eqref{corrdef}
 can be also carried out for the  initial states labeled by $(s/p,s)$ where  $\textbf{C}(0)$ has matrix elements
\begin{equation}\label{initial correlations}
[\boldsymbol{C}(0)]_{ln}=\delta_{ln}\left[1-\frac{1}{p}\sum _{j=0}^{p-1}\sum_{l'=0}^{s-1}e^{-2\pi i \frac{(l+l')j}{p}}\right];
\end{equation}
we will also briefly investigate such a possibility. 

\section{Light-cone velocities for the signal $\Delta_{ln}(t)$}
\label{sec4}
In this section we extract the light-cone velocity $v_{\max}$ for the connected two-point correlation function \eqref{corrdef} in the XY chain.
We divide our presentation in three subsections.  We first focus on XX chain ($\gamma=h=0$), where an exhaustive classification can be performed and then study quenches from
the N\'eel state ($p=2$) where also a complete picture emerges. Finally in the last subsection we examine quenches from states with $p>2$ where the calculation of the
light-cone velocities is considerably  more involved and show an example for the Ising chain.

\subsection{Free fermions $(J=-1, \gamma=0)$}
\label{secff}
\begin{figure} [t] 
\includegraphics[width=\columnwidth,angle =0]{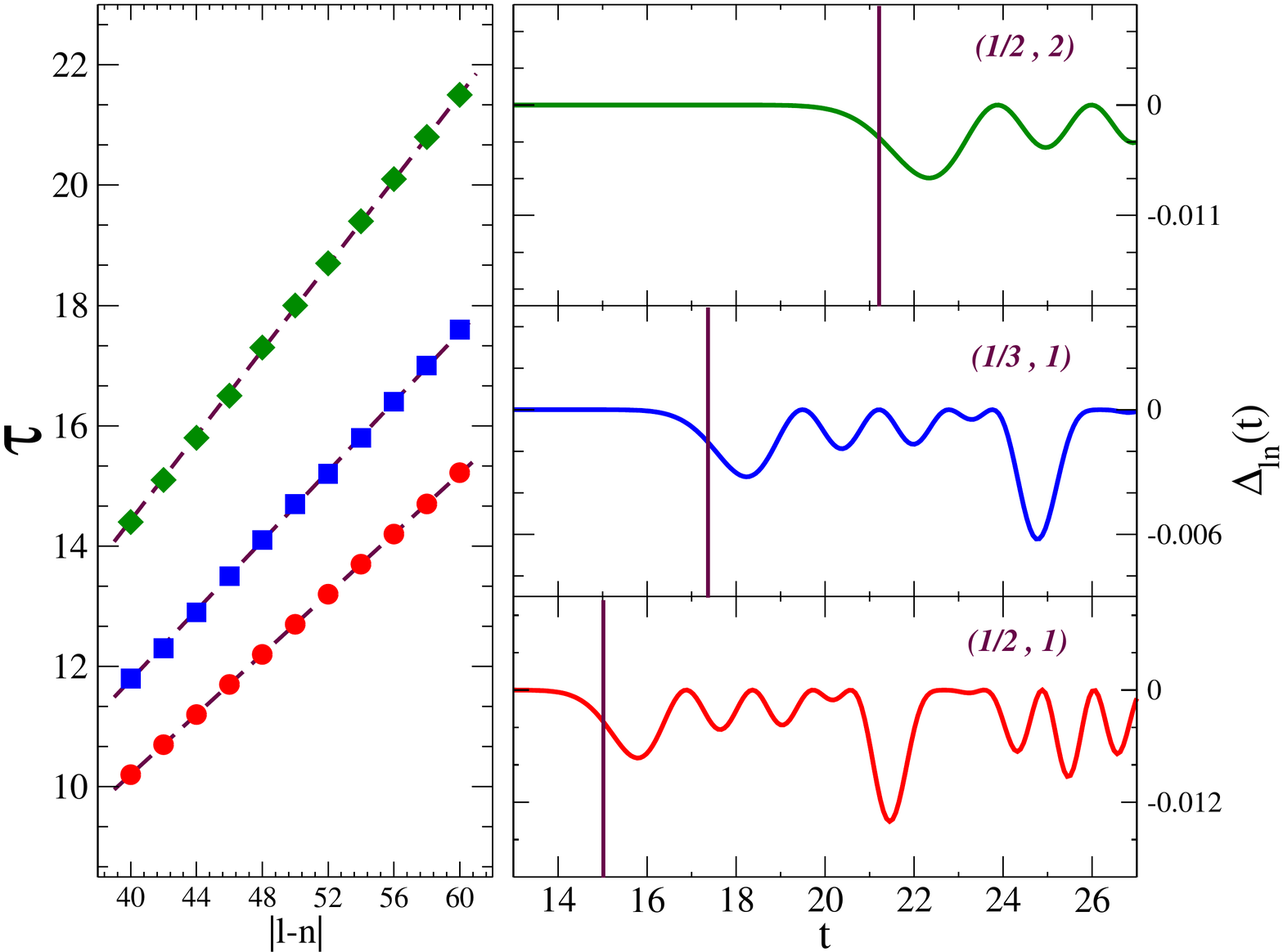}
\caption{(color online) On the right panel $\Delta_{ln}(t)$ for the XX chain ($\gamma=h=0$) and $J=-1$ with different initial states;  here we have $L=144$ and $|l-n|=60$. 
Vertical bars in correspondence of $\frac{|l-n|}{2v_{\max}}$. On the left panel we plotted the numerically estimated inflection points $\tau$ of the signals  as a function 
of the distances $|l-n|$. The points align on a straight line with slope given in Eq.~\eqref{XX max group velocity}.} 
\label{fig:XX}
\end{figure}
We consider preliminary the simple case of the XX chain ($\gamma=h=0$ and $J=-1$) where the fermion number is conserved and $\textbf{B}$ (and $\textbf{F}$) vanishes.
Then the matrix $\textbf{C}$ for the initial states labeled by $(1/p,1)$ can be rewritten as 
\begin{equation}
\label{fln_terms2 XX}
C_{ln}=\delta_{ln}-\frac{1}{Lp}\sum_{j=0}^{p-1}e^{-\frac{2\pi i n j}{p}}\sum_{k=1}^L e^{-i\left[\phi_{k}(l-n)+\left(\varepsilon_k-\varepsilon_{\frac{Lj}{p}-k}\right)t\right]},
\end{equation}
where $\varepsilon_k=-2\cos\phi_k$ and $\phi_k=\frac{2\pi k}{L}$ ($k=1,\dots,L$). Notice also that $v_g=2$.
For the sake of determining the light-cone velocity,
the last exponential in Eq.~\eqref{fln_terms2 XX} implies that
\begin{eqnarray}\label{XX effective dispersion relation XX}
\varepsilon^{\text{eff}}_{k,j}(p)\equiv\frac{1}{2}\left(\varepsilon_k-\varepsilon_{\frac{Lj}{p}-k} \right),
\end{eqnarray}
can be interpreted as an effective dispersion relation for the signal. The effective dispersion originates
from the discrete translational symmetry of $p$ lattice sites of the initial state that allows multiplets
of quasi-particles to be produced with momenta $k$ and $\frac{Lj}{p}-k$, $j=1,\dots,p$ 
(i.e. relaxing the condition of having only pairs with opposite momentum).
An effective group velocity can be defined from Eq.~\eqref{XX effective dispersion relation XX} as
\begin{eqnarray}\label{XX effective group velocity}
v^{\text{eff}}_{k,j}(p)=\sin\phi_k-\sin\left(\phi_k-\frac{2\pi j}{p}\right);
\end{eqnarray}
whose maximum value (over $j$ and $k$) is then the light-cone velocity
\begin{eqnarray}\label{XX max group velocity}
v_{\text{max}}= \begin{dcases*}
        2  & when $p$ is even,\\
        2\cos\left(\frac{\pi}{2p}\right) & when $p$ is odd.
        \end{dcases*}
\end{eqnarray}
 Eq.~\eqref{XX max group velocity} predicts quantitatively the light-cone velocity of the signal~\eqref{corrdef} for the free fermion case.
Note that the effective maximum group velocity occurs 
when $j=\frac{p}{2}$ and $j=\frac{p-1}{2}$ for even and odd $p$, respectively.
In particular, from Eq.~(\ref{XX max group velocity}) follows that the actual light-cone velocity is state-dependent and cannot be faster  than  the maximum group velocity $v_g$.
The prediction in Eq.~\eqref{XX max group velocity} is  in agreement with a numerical estimation of $v_{\max}$, obtained from the inflection points of the correlations
 $\Delta_{ln}$;  see Fig.~\ref{fig:XX} for examples with states labeled by $(1/2,1),~(1/3,1)$ and $(1/2,2)$. It is also interesting to observe that the absolute minimum visible in the second and third panel in  Fig.~\ref{fig:XX} is a finite size effect, indeed the envelope of $|\Delta_{ln}(t)|$ is monotonically decreasing in the limit $L\rightarrow\infty$ after reaching the first maximum.

A similar analysis for the initial states $(s/p,s)$, shows that 
\begin{equation}\label{fln_terms2 XX 2}
C_{ln}=\delta_{ln}-\frac{1}{Lp}\sum_{j=0}^{p-1}A_{js}~
e^{-\frac{2\pi i n j}{p}}\sum_{k=1}^L e^{-i\left[\phi_k(l-n)+2t\varepsilon^{\text{eff}}_{k,j}(p)\right]}
\end{equation}
where $A_{js}=\sum_{q=0}^{s-1}e^{-\frac{2\pi i j q}{p}}$. It is clear that as long as $A_{js}\not=0$ for the values of $j$
corresponding to the effective maximum group velocity, Eq.~(\ref{XX max group velocity}) remains valid.
However, one can verify that $A_{js}$  is actually zero in some cases.
For example, for the state labeled by $(1/2,2)$, we have $A_{22}=0$ and therefore the light-cone velocity
is obtained from the maximum over of Eq.~\eqref{XX effective group velocity} at $j=1$ and $p=4$; namely  $v_{\max}=\sqrt{2}$.
This is nicely confirmed in Fig.~\ref{fig:XX} (green curves). 
\subsection{Arbitrary values of $\gamma$ and $h$: Quenches from the N\'eel state $(p=2)$}
\label{sec3n}
We start our analysis of the light-cone velocity for arbitrary values of $\gamma$ and $h$ in the XY chain from the neatest case of a quench from the N\'eel
state; i.e. $p=2$ according to the notations of Sec.~\ref{sec2}.
First observe that, for arbitrary values of $\gamma$ and $h$, each
term in the sum over $j$ in Eqs.~\eqref{fln_terms2}-\eqref{cln_terms2} is associated to
a time propagation  with an effective dispersion 
\begin{equation}\label{XY effective dispersion relation generic}
\epsilon^{\text{eff}}_{k,j,\pm}(p)=\frac{1}{2}\left(\varepsilon_k\pm\varepsilon_{\frac{Lj}{p}-k}\right),~j=0,\dots,p-1,
\end{equation}
where it is understood that for $j=0$ we get back Eq.~\eqref{dispersion}. Moreover, also the first  line
in Eqs.~\eqref{fln_terms2}-\eqref{cln_terms2} contains a  state-independent contribution whose time evolution expands over the usual dispersion.
Therefore as long as $\gamma\not=0$, one should expect a state-independent
light-cone velocity $v_{\max}=v_g$ (the maximum group velocity) for all the functions~\eqref{flndef}-\eqref{clndef}. 
\begin{figure} [t] 
\includegraphics[width=\columnwidth,angle =0]{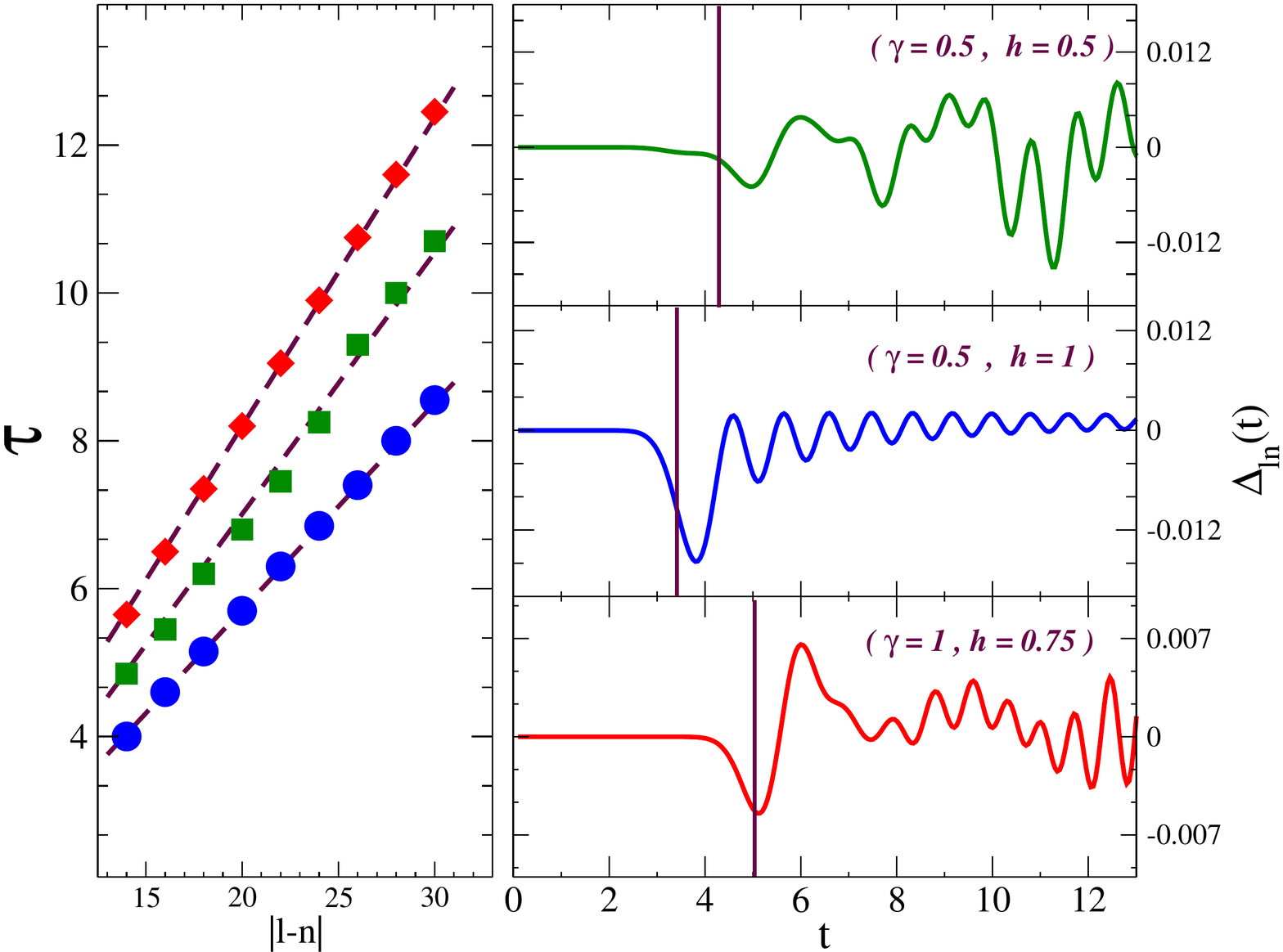}
\caption{(color online) On the right panel, $\Delta_{ln}(t)$ for the XY chain  and $J=1$ starting from the state $(1/2,1)$ with vertical bars
in correspondence of $\frac{|l-n|}{2v_{\max}}$;  here  $L=96$ and $|l-n|=12$. On the left panel we plotted the numerically estimated inflection points $\tau$ of the signals  as a function of the distances $|l-n|$. The points align on a straight line with slope obtained from Eq.~\eqref{vLC}.} 
\label{fig:XY1}
\end{figure}
However, as we already noticed at the end of Sec.~\ref{sec3}, for $p=2$  a propagation with the maximum group velocity $v_g$ is
forbidden by Eq.~\eqref{ T22 and T12 equality}.
For arbitrary values of the parameters $\gamma$ and $h$, the effective maximum group velocity (i.e. the light-cone velocity)  can be then obtained
from Eq.~\eqref{XY effective dispersion relation generic} as
\begin{equation}
\label{vLC}
v_{\text{max}}=\max_{j\not=0,\pm,k}\frac{d\varepsilon^{\text{eff}}_{k,j\not=0,\pm}(p)}{dk}.
\end{equation}
Actually if $p=2$, only $j=1$ is allowed in the formula above and as an immediate consequence $v_{\max}<v_{g}$;  Eq.~\eqref{vLC} predicts the light-cone velocity
of the signal~\eqref{corrdef} after a a quench from the N\'eel state in the XY chain.

The physical interpretation is similar to the free fermion case: here the initial state allows only pairs of quasi-particles to propagate with momenta
$k$ and $\frac{L}{2}-k$. Notice also that the velocities of the pair
are now different since $|v_k|\not=|v_{\frac{L}{2}-k}|$ for $v_k$ as in Eq.~\eqref{group velocity}; this asymmetry  diminishes the light-cone velocity. 
In the   Ising chain ($\gamma=1$)  $v_{\max}$ is obtained selecting the negative sign in
Eq.~\eqref{vLC}, and $\phi_k$ as close as possible to $\frac{\pi}{2}$. The explicit value is 
\begin{equation}\label{Ising MGV}
v_{\max}=\frac{2Jh}{\sqrt{1+h^2}}.
\end{equation}
Similarly, in the regions of the parameters $h=1$ and $\gamma<\gamma^*$, one finds
\begin{equation}\label{Ising MGV2}
v_{\max}=\frac{2J}{\sqrt{1+\gamma^2}};
\end{equation}
where $\gamma^*$ is the solution of $\gamma^*=\frac{1}{\sqrt{1+(\gamma^*)^2}}$.
In general,  there is not a closed formula for $v_{\max}$; see however the first and second panel in Fig.~\ref{fig:velocities} for numerical estimations
based on Eq.~\eqref{vLC}.
In Fig.~\ref{fig:XY1}, we plot $\Delta_{ln}(t)$ for different values of $\gamma$ and $h$ for quenches from the
initial configuration $(1/2,1)$.  The light-cone velocities estimated from the inflection points 
are in  agreement with Eqs.~\eqref{Ising MGV}-\eqref{Ising MGV2} and more generally with Eq.~\eqref{vLC}.
\subsection{Arbitrary values of $\gamma$ and $h$: Quenches from states with $p>2$}
\begin{figure} [t] 
\includegraphics[width=\columnwidth,angle =0]{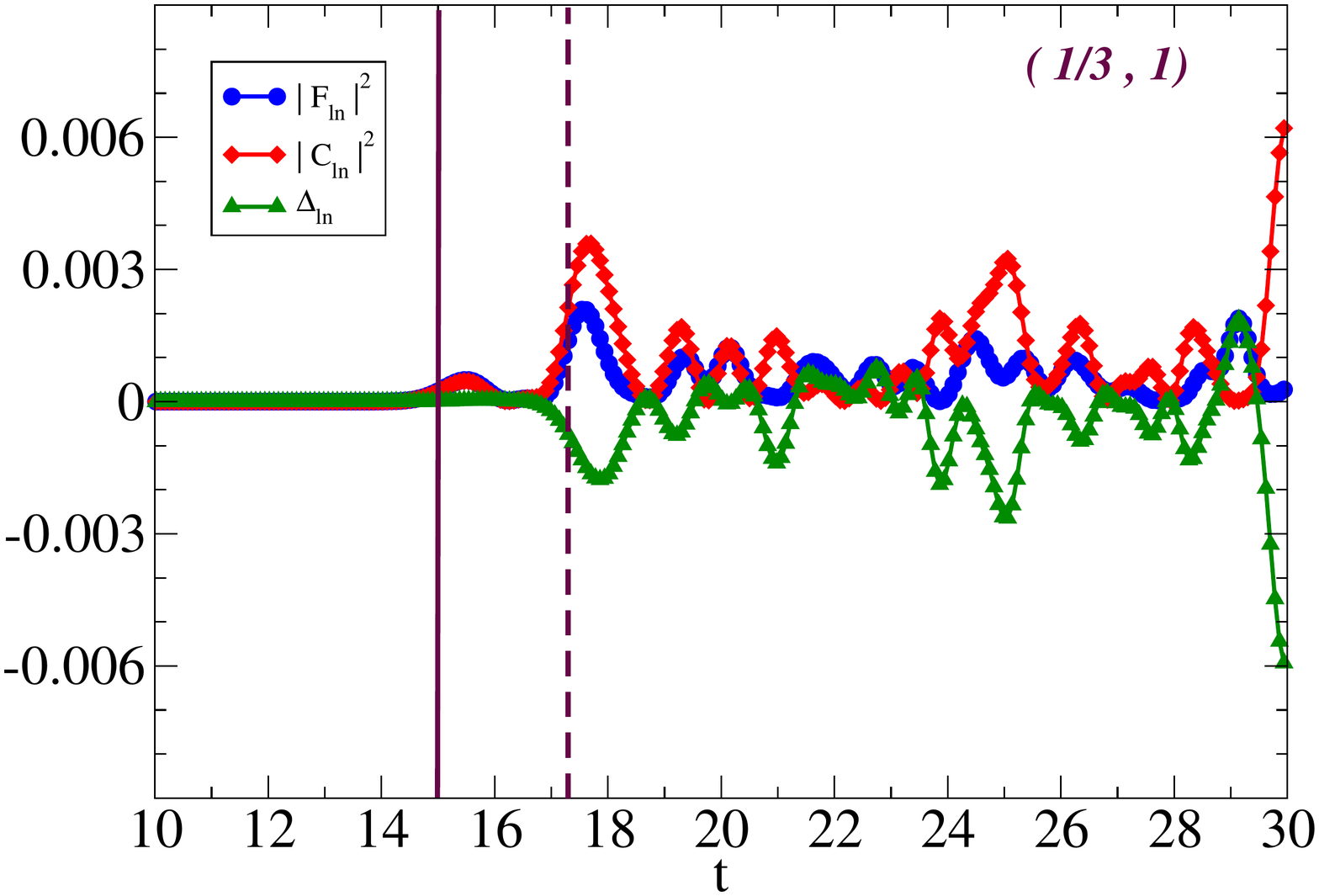}
\includegraphics[width=\columnwidth,angle =0]{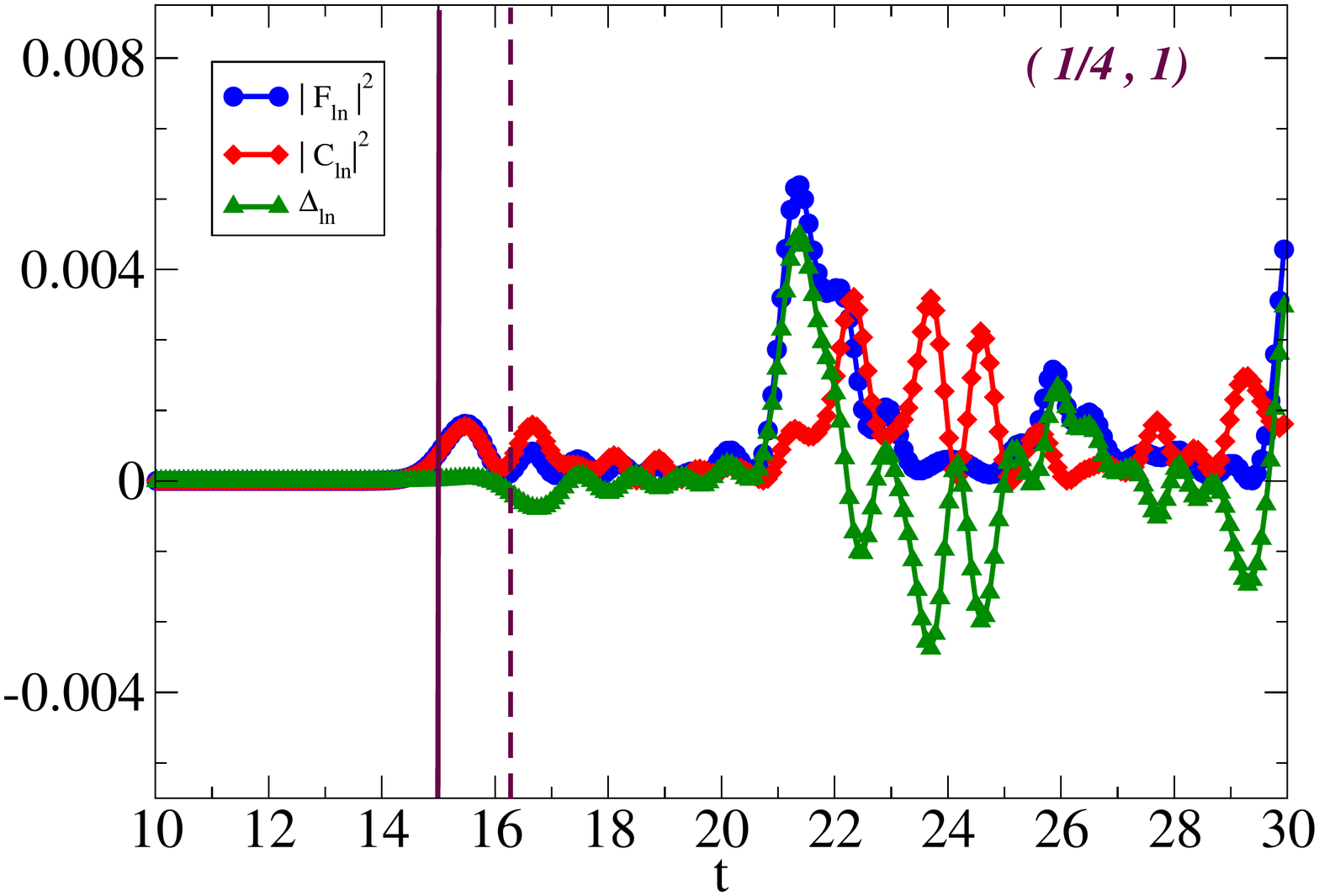}
\caption{(color online) $|C_{ln}|^{2}$, $|F_{ln}|^{2}$, and $\Delta_{ln}(t)$ for the XY chain with $J=1$, $\gamma=1$ and $h=1$ (critical Ising chain).
 The continuous vertical lines intersecting the red and blue curves
are indicating the inflection points $\tau$. The dashed vertical line is in correspondence of the inflection point of $\Delta_{ln}$ (green curve). It can be seen that $|F_{ln}|$ and $|C_{ln}|$ propagates faster than $\Delta_{ln}$. Here $L=144$ and $|l-n|=60$.} 
\label{fig:XY2}
\end{figure}

For $p>2$, the determination of $v_{\max}$ is considerably more involved and we cannot provide explicit formulas covering all the cases.
To understand the source of new subtleties  let us  focus first on the critical Ising chain.
Since for $p>2$ the  terms in the first line of Eqs.~(\ref{fln_terms2})-(\ref{cln_terms2}) are not canceled,  we should expect 
that the $\textbf{C}$ and $\textbf{F}$  matrix elements will propagate with a state-independent velocity $v_g$.
This is indeed correct, see for instance the red  and blue curves in the two panels of Fig.~\ref{fig:XY2} with initial states $p=3$ and $p=4$. 
However, when combined into  $\Delta_{ln}$, the two signals  almost exactly cancel around the first maximum,
leaving a light-cone velocity slower than $v_g$. The latter can be still calculated as in case $p=2$ from Eq.~\eqref{vLC}.
See the green curve in Fig.~\ref{fig:XY2} for an illustration. This result holds also for any $h\leq 1$ (ferromagnetic phase)
as we support analytically  in the Sec.~\ref{app}. For $h>1$ (paramagnetic phase), the same
Sec.~\ref{app}  shows instead that the correlation function $\Delta_{ln}$ propagates  with the maximum group velocity $v_g$.
Summarizing for $p>2$, in the Ising chain the signal in Eq.~\eqref{corrdef} propagates with a light cone velocity given by Eq.~\eqref{vLC} for $h\leq 1$ and by the maximum group
velocity $v_g$ for $h>1$.
This difference in the light-cone velocity between the two phases is hard to spot numerically, since  the state-independent term is of order $1/h^2$
and the difference between  Eq.~\eqref{vLC} and $v_g$ drops to zero fast as $h$ increases.

We studied the correlation function $\Delta_{ln}$ for several different values of the parameters $\gamma$ and $h$. For the initial states $(\frac{1}{p},1)$
with $p>2$ we found numerically,  similarly to the Ising chain,  $v_{\max}$ always to be given by Eq.~\eqref{vLC} or $v_g$.
However a clear pattern does not emerge from the numerical analysis and to distinguish between the two cases one should resort to the method described
in  Sec.~\ref{app}.
For the configuration $(1/2,2)$ with $p=4$ again the terms that are independent from the initial states cancel out explicitly. For instance in the Ising chain   the light-cone velocity is
\begin{eqnarray}\label{Ising MGV p4}
v_{\text{max}}=\frac{\sqrt{2}Jh}{\sqrt{1-\sqrt{2}h+h^2}};
\end{eqnarray}
%
%
%
a result that can be verified numerically from the inflection points. Similar arguments are also valid for
all the configurations with $(s/p,s)$.

\section{Airy scaling of correlation functions in infinite volume}
\label{app}
In this section we describe how it is possible to obtain accurate approximations of the free fermionic correlation functions
near the edge of the light cone in terms of combinations of Airy functions.
The appearance of the Airy function at the boundary of the light-cone is not a novelty in the quench context, see~\cite{CAiry, B2016} and also \cite{current, Arctic}. It is
actually a direct consequence of stationary phase approximation in presence of coalescing stationary points;
for more details and possible sources of violation of this approximation, see the comments below  Eq.~\eqref{statphase}.

We start by determining
the infinite volume limit ($L\rightarrow\infty$) of the fermionic propagator and recast the results in Sec.~\ref{sec3} directly in Fourier space.  
We consider then fermionic operators
$c_l$ and $c_n^{\dagger}$ with $\{c^{\dagger}_l,c_{n}\}=\delta_{ln}$, $\{c_l,c_{n}\}=\{c^{\dagger}_l,c^{\dagger}_{n}\}=0$ defined for any
$l,n\in\mathbb Z$. The convention for the Fourier transforms are
\begin{equation}
\label{Fourier}
 c_l=\int_{-\pi}^{\pi}\frac{dk}{\sqrt{2\pi}}~e^{ikl}c(k),~c(k)=\frac{1}{\sqrt{2\pi}}\sum_{l\in\mathbb Z} e^{-ikl}c_l,
\end{equation}
from which follows that $\{c^{\dagger}(k),c(k')\}=\delta(k-k')$
and $\{c(k),c(k')\}=\{c^{\dagger}(k),c^{\dagger}(k')\}=0$. To be definite we only consider  initial configurations labeled by
the pair $(1/p,1)$.
In the XY chain, the time evolved operators $c(-k,t)$ and $c^{\dagger}(k,t)$ are linearly related
to  the correspondent operators at time zero. Indeed the Bogoliubov rotation that diagonalizes the XY chain is~\cite{LSM}
\begin{equation}
\label{Bog}
\begin{pmatrix}
b^{\dagger}(k)\\
b(-k)
 \end{pmatrix}=\overbrace{\begin{pmatrix} \cos(\theta(k)/2)  & -i\sin(\theta(k)/2)  \\
 -i\sin(\theta(k)/2)& \cos(\theta(k)/2)  \end{pmatrix}}^{\mathcal R(k)} \begin{pmatrix}
c^{\dagger}(k)\\
c(-k)
 \end{pmatrix}
 \end{equation}
 with $\cos\theta(k)=\frac{2J(\cos(k)-h/J)}{\varepsilon(k)}$,
 $\sin\theta(k)=\frac{2J\gamma\sin(k)}{\varepsilon(k)}$ and
 $\varepsilon(k)=2J\sqrt{(\cos(k)-h/J)^2+\gamma^2\sin^2(k)}$. The Bogoliubov fermions $b(k)$
 and $b^{\dagger}(-k)$ have simple time evolution
 \begin{equation}
 \begin{pmatrix}
b^{\dagger}(k,t)\\
b(-k,t)
 \end{pmatrix}=\overbrace{\begin{pmatrix} e^{i\varepsilon(k)t}  & 0  \\
 0 & e^{-i\varepsilon(k)t} \end{pmatrix}}^{\mathcal U(k,t)} \begin{pmatrix}
b^{\dagger}(k)\\
b(-k),
 \end{pmatrix}
 \end{equation}
and therefore we obtain 
\begin{equation}
\label{time_ev}
 \begin{pmatrix}
c^{\dagger}(k,t)\\
c(-k,t)
 \end{pmatrix}=\overbrace{\mathcal{R}^{\dagger}(k)\mathcal{U}(k,t)\mathcal{R}(k)}^{\tilde{T}(k,t)}\begin{pmatrix}
c^{\dagger}(k)\\
c(-k)
 \end{pmatrix}.
\end{equation}
The matrix elements of $\tilde{T}$  are
\begin{align}
\label{T11fourier}
&\tilde{T}_{11}(k,t)=\cos(\varepsilon(k)t)+i\cos(\theta(k))
\sin(\varepsilon(k)t)\\
\label{T12fourier}
&\tilde{T}_{12}(k,t)=
\sin(\varepsilon(k)t)\sin(\theta(k)).
\end{align}
and $\tilde{T}_{11}(k,t)=[\tilde{T}_{22}^*(k,t)]$,
$\tilde{T}_{21}(k,t)=-\tilde{T}_{12}(k,t)$. From unitarity it follows moreover
$|\tilde{T}_{11}|^2+\tilde{T}_{12}^2=1$
and $\tilde{T}_{11}^*\tilde{T}_{12}+\tilde{T}_{21}^*\tilde{T}_{22}=0$. Fermionic correlation functions are double integrals in Fourier space. For instance let
us denote by $f^{\alpha}$ a fermionic operator with $f^{+}\equiv c$ and 
$f^{-}\equiv c^{\dagger}$, from the definition of the Fourier
transform~\eqref{Fourier} we have
\begin{equation}
\label{double_int}
 \langle f^{\alpha}_l(t)f^{\beta}_n(t)\rangle=\int\int\frac{dk dk'}{2\pi}
 e^{i\alpha kl +i\beta k'n}\langle f^{\alpha}(k,t)f^{\beta}(k',t)\rangle,
\end{equation}
with integrals in the domain $k,k'\in[-\pi,\pi]$.
The time evolution of the matrix element in~\eqref{double_int} is obtained from
Eq.~\eqref{time_ev} as a linear combination of four matrix elements of the same
type at time $t=0$.
Among those four the only non-trivial  on the class of initial states
we are considering is $g(k,k')=\langle c^{\dagger}(k) c(k')\rangle$.
Notice indeed that
$\langle c^{\dagger}(k)c^{\dagger}(k')\rangle=
\langle c^{\dagger}(k)c^{\dagger}(k') \rangle=0$ and $\langle c(k)c^{\dagger}(k') \rangle$ can be obtained from the anticommutation relations. 
For our initial state $(\frac{1}{p},1)$ the function $g$ is given by
\begin{equation}
\label{state}
 g(k,k')=\frac{1}{2\pi}\sum_{n\in\mathbb Z}e^{inp(k-k')}=
 \frac{1}{p}\sum_{j=0}^{p-1}\delta\left(k-k'-\frac{2\pi j}{p}\right).
\end{equation}
It is then straightforward to derive integral
representations for the correlators $F_{ln}(t)$ and $C_{ln}(t)$ that we write as
\begin{equation}
\label{FC}
 F_{ln}(t)=\sum_{j=0}^{p-1}F_{ln}^{j}(t)~~\text{and}~~C_{ln}(t)=\sum_{j=0}^{p-1}C_{ln}^{j}(t).
\end{equation}
Explicit expressions for the functions $F_{ln}^j(t)$ and $C_{ln}^j(t)$ are
 \begin{multline}
 \label{Ffourier0}
F_{ln}^{0}(t)=\int\frac{dk}{2\pi} e^{-ik(l-n)}\left[\tilde{T}_{12}(k,t)\tilde{T}_{11}(-k,t)+\right.\\
\left.\frac{1}{p}(\tilde{T}_{11}(k,t)\tilde{T}_{12}(-k,t)-
\tilde{T}_{12}(k,t)\tilde{T}_{11}(-k,t))\right],
\end{multline}
\begin{multline}
\label{Ffouriernt0}
F_{ln}^{j\not=0}(t)=\frac{e^{-\frac{2\pi i n j}{p}}}{p}\int\frac{dk}{2\pi} e^{-ik(l-n)}
 \left[\tilde{T}_{11}(k,t)\times\right.\\
\left.\times\tilde{T}_{12}\Bigl(-k+\frac{2\pi j}{p},t\Bigr)-\tilde{T}_{12}(k,t)\tilde{T}_{11}\Bigl(-k+\frac{2\pi j}{p},t\Bigr)\right];
\end{multline}
 and analogously
 \begin{multline}
 \label{Cfourier0}
  C^0_{ln}(t)=\int\frac{dk}{2\pi} e^{ik(l-n)}\left[\tilde{T}_{22}(-k,t)\tilde{T}_{11}(k,t)\left(1-\frac{1}{p}\right)+\right.\\
  \left.\frac{1}{p}\tilde{T}_{12}(k,t)^2\right],
 \end{multline}
 \begin{multline}
 \label{Cfouriernot0}
 C^{j\not=0}_{ln}(t)=\frac{e^{-\frac{2\pi i nj}{p}}}{p}\int\frac{dk}{2\pi}e^{ik(l-n)}\left[\tilde{T}_{21}(-k,t)\times\right.\\
 \left.\times\tilde{T}_{12}\Bigl(k+\frac{2\pi j}{p},t\Bigl)
-\tilde{T}_{22}(-k,t)\tilde{T}_{11}\Bigl(k+\frac{2\pi j}{p},t\Bigr)\right].
 \end{multline}
 Notice that the integrals in Eqs.~\eqref{Ffourier0} and \eqref{Cfourier0} are of course the $L\rightarrow\infty$ limit of Eqs.~\eqref{fln_terms2}-\eqref{cln_terms2}. Each function $F_{ln}^j(t)$
and $C_{ln}^j(t)$ 
describes a time propagation with a velocity that can be derived from the
effective dispersion relation
\begin{equation}
\label{effective_d}
 \varepsilon^{\text{eff}}_{j,\pm}(k,p)=\frac{1}{2}\left[\varepsilon(k)\pm\varepsilon
 \left(-k+\frac{2\pi j}{p}\right)\right],~j=0,\dots,p-1,
\end{equation}
that is Eq.~\eqref{XY effective dispersion relation generic} in the limit $L\rightarrow\infty$. As in Sec.~\ref{sec3},
the prediction of the light-cone velocity follows from calculating
the maximum effective group velocities obtained from Eq.~\eqref{effective_d}. In particular, it can be easily verified
(see Eqs.~\eqref{Ffourier0} and \eqref{Cfourier0} in particular) that for
$p=2$, the maximum effective group velocity is always smaller than $v_g$ obtained from Eq.~\eqref{group velocity}.

As shown in Fig.~\ref{fig:XY2}, the numerics indicates that 
a cancellation of the fastest $j=0$ contributions in Eq.~\eqref{effective_d} appears also for $p\not=2$ in the critical Ising chain.
This observation extends to the whole ferromagnetic phase $h\leq 1$. It can be understood analytically 
comparing the  behaviors  near their inflection points of $|F^0_{ln}(t)|$ and $|C^0_{ln}(t)|$ in \eqref{FC} and showing that they are the same. 
At the edge of the light-cone those functions can indeed be approximated by an Airy function with increasing accuracy as $t\rightarrow\infty$ as we now discuss. 

Consider an integral of the form 
\begin{equation}
\label{Airy_int}
 I(x,t)=\int_{-\pi}^{\pi}\frac{dk}{2\pi} H(k)e^{2it\varepsilon(k)-ikx}
\end{equation}
where we assume $x>0$ and $H(k)$ a function with support into  $[-\pi,\pi]$. We consider the asymptotic of such an integral 
for large $t$, keeping fixed the ratio $x/t$. The boundary of the light-cone is identified by the condition that  the solution $k_{\max}$ of the stationary phase equation 
\begin{equation}
\label{statphase}
\varepsilon'(k_{\max})=\frac{x}{2t},
\end{equation}
corresponds to a point of maximum of the function $\varepsilon'(k)$. There are then two possibilities: either $\varepsilon''(k_{\max})=0$ or $\varepsilon''(k_{\max})\not =0$. The latter case can happen for instance if  $k_{\max}$ lies at the boundary of the integration domain. The Airy scaling is obtained when $\varepsilon''(k_{\text{max}})=0$ and in a Taylor expansion of the phase near $k=k_{\max}$  the third order term is not zero (for instance~\cite{B2016, current, Arctic}). If $H(k_{\max})\not=0$, it is easy to obtain the following approximation of the integral $I(x,t)$
near the boundary of the light-cone
\begin{equation}
\label{Airy_approx}
 I(x,t)\simeq\frac{e^{2it\varepsilon(k_{\max})-ik_{\max}x}H(k_{\max})}{[-t \varepsilon'''(k_{\max})]^{1/3}}
 \text{Ai}(-X),
\end{equation}
being  $X=\frac{2\varepsilon'(k_{\max})t-x}{[-t\varepsilon'''(k_{\max})]^{1/3}}$ and $\text{Ai}(x)=\int_{-\infty}^{\infty}\frac{dq}{2\pi}e^{iqx+\frac{iq^3}{3}}$.
It should be noticed that~\eqref{Airy_approx} is determined in the limit $t\rightarrow\infty$  but it gives a
 fairly good approximation of the integral as long as 
 \begin{equation}
 |2\varepsilon'(k_{\max})t-x|\ll [-t \varepsilon'''(k_{\max})]^{1/3}.
 \end{equation}
 
Let us now pass to illustrate a concrete application of Eq.~\eqref{Airy_approx} to the Ising chain where we fixed  $J=1/2$ and $\gamma=1$.
The functions
$F^0_{ln}(t)$ and $C^0_{x}(t)$ can be
written as 
\begin{align}
\label{F0full}
 & F^0_{ln}(t)=H_0+\sum_{\sigma=\pm}\int_{-\pi}^{\pi}\frac{dk}{2\pi}H_{\sigma}(k)e^{ 2 i\sigma\varepsilon(k)t-ik(l-n)},\\
 \label{C0full}
 & C^0_{ln}(t)=K_0+\sum_{\sigma=\pm}\int_{-\pi}^{\pi}\frac{dk}{2\pi}K_{\sigma}(k)e^{ 2 i\sigma\varepsilon(k)t-ik(l-n)};
\end{align}
where $H_{\pm}$ and $K_{\pm}$ are obtained expanding the integrands in \eqref{Ffourier0} and \eqref{Cfourier0} and we also used $C^{0}_{ln}(t)=C^{0}_{nl}(t)$.
The constant terms $H_0$ and $K_0$ can be also determined explicitly from the residue theorem and they are given by
\begin{align}
& H_0=K_0=-\frac{h^{x-2}(h^2-1)(p-2)}{8p},\quad h\leq 1;\\
& H_0=-K_0=-\frac{h^{-x+2}(h^2-1)(p-2)}{8p},\quad h>1,
\end{align}
where we defined $x\equiv(l-n)\geq 2$.
They are then  vanishing at $h=1$ or are exponentially small with the distance $x$ when $h\not=1$. Since we will  consider only large values of $x$, it turns out that they are negligible. 

 We focus first on the case $h<1$. Here we have $k_{\max}=\arccos(h)$,  $\varepsilon'(k_{\max})\equiv v_g=h$, $\varepsilon''(k_{\max})=0$, $\varepsilon'''(k_{\max})=-h$; using
 \eqref{Airy_approx} we obtain for $t\rightarrow\infty$ the light-cone approximations
 \begin{align}
 \label{Airy1}
 & F^{0}_{ln}(t)\simeq\frac{2H_+(k_{\max})\cos[2t h\phi(h)]}{(ht)^{1/3}}\text{Ai}(X)\\
 \label{Airy2}
 & C^0_{ln}(t)\simeq\frac{2K_+(k_{\max})\cos[2th\phi(h)]}{(ht)^{1/3}}\text{Ai}(X).
 \end{align}
 where $\phi(h)=\sqrt{1/h^2-1}-\arccos(h)$.
It turns out that $H_{+}(k_{\max})=H_{-}(-k_{\max})=-iK_{+}(k_{\max})=-iK_{-}(-k_{\max})$ and 
\begin{equation}
\label{H+1}
H_{+}(k_{max})=-i\frac{p-2}{4p},
\end{equation}
Therefore $|F^0_{ln}|^2$ and $|C^0_{ln}|^2$ have the same approximation in terms of an Airy function and they cancel out when calculating $\Delta_{ln}$.

We now pass to discuss the case $h>1$. Here we have $k_{\max}=\arccos(1/h)$,  $\varepsilon'(k_{\max})\equiv v_g=1$, $\varepsilon''(k_{\max})=0$,~$\varepsilon'''(k_{\max})=-1$. A similar expansion gives the approximations
\begin{align}
\label{Airy1a}
 \hspace*{-0.5cm}&F^{0}_{ln}(t)\simeq\frac{e^{2it\phi(h^{-1})}
 H_+(k_{\max})+e^{-2it\phi(h^{-1})}H_{-}(-k_{\max})}{t^{1/3}}\text{Ai}(X)\\
 \label{Airy2a}
 &C^0_{ln}(t)\simeq\frac{2K_+(k_{\max})
 \cos[2t\phi(h^{-1})]}{t^{1/3}}\text{Ai}(X),
\end{align}
where now $K_+(k_{\max})=K_-(-k_{\max})$ and
\begin{align}
\label{H+2}
&H_{\pm}(\pm k_{\max})=-i\frac{(h\mp\sqrt{h^2-1})(p-2)}{4h^2p}\\
\label{K+2}
&K_{+}(k_{\max})=\frac{p-2}{4h^2p}.
\end{align}
There is no more a cancellation of the two terms as for  $h<1$. The absence of a clear peak at $\tau=\frac{l-n}{2}$ ($l\gg n$) in the observable $\Delta_{ln}(t)$ that
emerges in the numerics can be however understood as a combination of three effects. For large $h\gg 1$, the values at the inclination point of $|F_{ln}^0|^2$ and $|C_{ln}^0|^2$ are suppressed by a factor $1/h^2$ and $1/h^4$ respectively and moreover the difference between Eq.~\eqref{vLC} and $v_g$ is small. On the other hand, as  $h\rightarrow 1$,  an exact cancellation between $|F_{ln}^0|$ and $|C_{ln}^0|$ must happen. 

For $h=1$, the two distinct stationary points  $\pm k_{\max}$ actually merge at $k=0$ and the asymptotic expansion involves only one term. One has $v_g=1$ and $\varepsilon''(0^+)=0$,
 $\varepsilon'''(0^+)=-1/4$; leading to the final result
 \begin{equation}
 \label{Airy3}
 F^{0}_{ln}(t)\simeq-i\frac{p-2}{2p (2t)^{1/3}}\text{Ai}(X),\quad 
  C^0_{ln}(t)=-iF^{0}_{ln}(\tau),
 \end{equation}
that once again shows the cancellation of the fastest particle contribution for arbitrary values of $p$.
Although the argument applies
 for large $t$, we believe that it  furnishes a satisfactory explanation of the numerical results  presented in Fig.~\ref{fig:XY2}. Finally notice  that the limits $h\rightarrow 1^{\pm}$, although producing the same result, do not commute with the asymptotic expansion. 

\begin{figure} [t] 
\includegraphics[width=0.45\textwidth,angle =0]{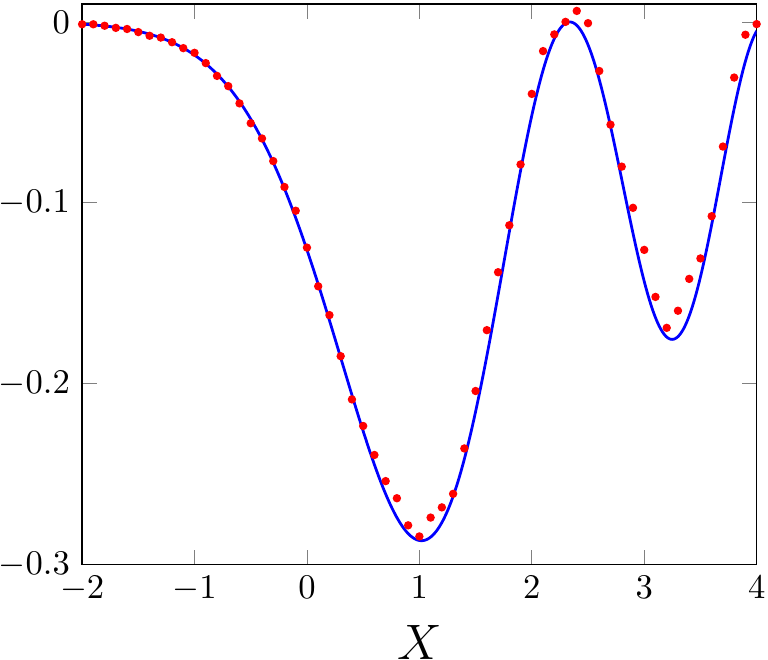}
\caption{The continuous blue curve is the function $-\text{Ai}^2(-X)$. The points represent numerical evaluations
 of $(4t)^{2/3}\Delta_{ln}(t)$ for a quench from the N\'eel state ($p=2$) in the critical Ising chain ($h=\gamma=1$ and $J=1/2$).
  Here $|l-n|=2v_{\text{max}}t-[-te'''(k_{\max})]^{1/3}X$ and $t=500$.}
  \label{fig_airy}
\end{figure}

Analogous approximations  could be calculated for all the functions $F^{j}_{ln}(t)$ and $C^{j}_{ln}(t)$ with the same technique.
As a last example, we demonstrate the validity of the Airy-approximation,
through the neatest example of a quench from the N\'eel state ($p=2$) in the critical Ising chain. In this case the
integrals $F^0_{ln}(t)$ and $C_{ln}^0(t)$ vanish for any time. For $t\rightarrow\infty$ and $x/t$ finite,
being $x\equiv l-n>0$;  we obtain
\begin{equation}
\label{Airy_delta}
(4t)^{2/3}\Delta_{ln}(t)\simeq-\text{Ai}^2(-X),\quad t\gg 1.
\end{equation}
In \eqref{Airy_delta}, $X=\frac{2v_{\text{max}}t-x}{[-te'''(k_{\max})]^{1/3}}\in\mathbb R$, $k_{\max}=\pi/2$ and according
to  (\ref{effective_d})
\begin{align}
& e(k)\equiv\varepsilon_{1,-}^{\text{eff}}(k,2)=\frac{1}{2}[\varepsilon(k)-\varepsilon(-k+\pi)]\\
& v_{\text{max}}=\max_{k\in[-\pi,\pi]}\frac{de(k)}{dk}=\frac{1}{\sqrt{2}}.
\end{align}
The value of $v_{\max}$ is of course the same as in Eq.~\eqref{vLC} taking $J=1/2$.
A comparison of the stationary phase approximation in Eq.~\eqref{Airy_delta} with a numerical evaluation of the correlation function $\Delta_{ln}(t)$ is presented in Fig.~\ref{fig_airy}. It is finally worth to remark that
for $X=0$ (i.e. $t=\frac{|l-n|}{2v_{\text{max}}}$),
 Eq.~\eqref{Airy_delta} is within the $10\%$ from the exact value already for $t=50$. Notice also that $\text{Ai}^{\prime\prime}(0)=0$ which explains why for large times
 the inflection point of the signal is
 close to $\frac{|l-n|}{2v_{\text{max}}}$. 
\section{Evolution of the entanglement entropy}
\label{sec5}
\begin{figure}[t] 
\begin{tikzpicture}[scale=0.65]
\node at (0,0) {\includegraphics[width=1\columnwidth,angle =0]{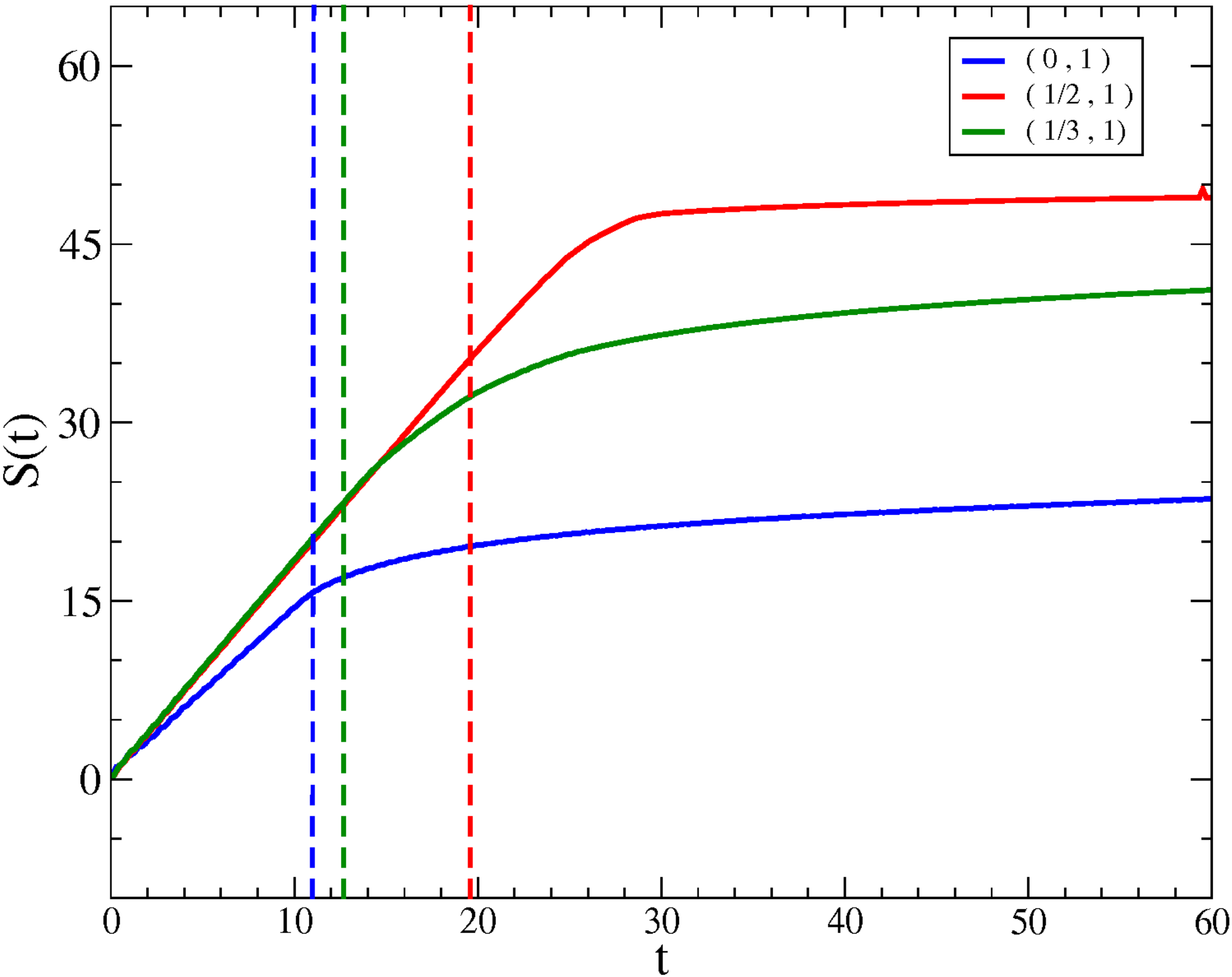}};
\node at (-5, -5.5) {\large{(a)}};
\node at (0,-11.5) {\includegraphics[width=1\columnwidth,angle =0]{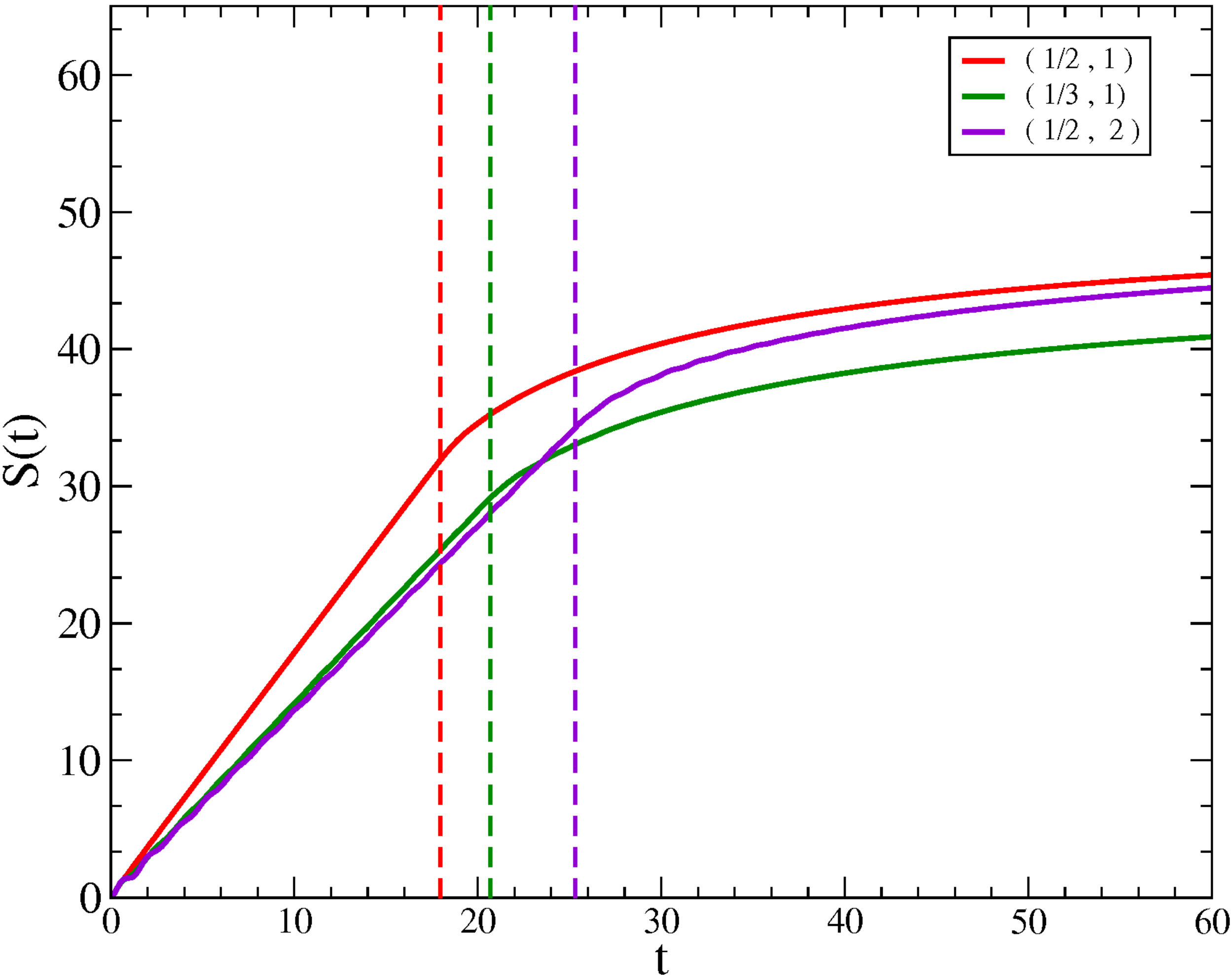}};
\node at (-5, -5.5-11.5) {\large{(b)}};
\end{tikzpicture}
\caption{(color online) The evolution of the entanglement entropy
in the XY chain with  system size $L=600$ and $l=72$. The vertical dashed lines are in correspondence
of  $t=\frac{l}{2v_{\max}}$, being $v_{\max}$  the light-cone velocity 
for the correlation functions obtained in Sec.~\ref{sec4}. 
(a) $S(t)$ at $\gamma=2$ and $h=1.5$. The blue curve corresponds to time evolution from $p=1$
(i.e. a fully polarized
initial state) and the green one to $p=3$. Both are compatible
with $\tau_s=\frac{l}{2v_{\max}}$, notice that for $p=1$ $v_{\max}=v_g$. The red curve
corresponds instead to $p=2$ and shows  that
in such a case the saturation time is larger than $\frac{l}{2v_{\max}}$.
(b) $S(t)$ in  the free fermion case $\gamma=h=0$ and $J=-1$, here the saturation time is
in  agreement with the conjecture $\tau_s=\frac{l}{2 v_{\max}}$.}
\label{fig:ent}
\end{figure}
In this final section before the conclusions, we present numerical results for the time evolution of the entanglement entropy
of a subsystem of size $l$ for the different initial states discussed in Sec.~\ref{sec2}. 
Based on a quasi-particle picture~\cite{CC2006, CC2007}, we expect the entanglement entropy, denoted by  $S(t)$, to grow linearly in time
up to $\tau_s$ which is approximately $\frac{l}{2v_g}$. 
After $\tau_s$, which we will call the saturation time,  the entanglement entropy converges~\cite{AC2017} to the von Neumann entropy of the stationary state
~\cite{CE2013, foot2}. In the 
numerics $\tau_s$ is obtained as the earliest  time where
the second derivative of the signal changes.

One can calculate the entanglement entropy from the correlation functions as follows~\cite{Peschela}
\begin{equation}\label{ent2}
S=- {\rm Tr}\left[\frac{1+\boldsymbol{\Gamma}}{2}\ln\left(\frac{1+\boldsymbol{\Gamma}}{2}\right)\right],
\end{equation}
where $\boldsymbol{\Gamma}$ is a $2l\times 2l$ block matrix which can be written as 
\begin{align}
\nonumber
\boldsymbol{\Gamma}_{mn}&=\left\langle {{a_{m}^{x}}\choose{a_{m}^{y}}}(a_{n}^{x}~ a_{n}^{y}) \right\rangle
-\delta_{mn}\mathbf{1}_{2\times 2}\\ \label{gamma1}
&= \begin{pmatrix}
\langle a_{m}^{x}a_{n}^{x} \rangle -\delta_{mn} & \langle a_{m}^{x}a_{n}^{y} \rangle \\
\langle a_{m}^{y}a_{n}^{x} \rangle  &\langle a_{m}^{y}a_{n}^{y} \rangle -\delta_{mn}\\
  \end{pmatrix}.
\end{align} 
Here $a_{m}^{x}=c_{m}^{\dagger}+c_{m}$ and $a_{m}^{y}=i(c_{m}-c_{m}^{\dagger})$ and the indexes $m,n$ belong to the one-dimensional
subsystem of size $l$.
One can easily find all the different elements of the matrix $\boldsymbol{\Gamma}$ as,

\begin{eqnarray}
\boldsymbol{\Gamma}_{11}&=&\boldsymbol{F}+\boldsymbol{F}^{\dagger}+\boldsymbol{C}-\boldsymbol{C}^{T},\\
\boldsymbol{\Gamma}_{12}&=&i(\mathbf{1}-\boldsymbol{C}-\boldsymbol{C}^{T}-\boldsymbol{F}+\boldsymbol{F}^{\dagger}),  \\
\boldsymbol{\Gamma}_{21}&=&-i(\mathbf{1}-\boldsymbol{C}-\boldsymbol{C}^{T}+\boldsymbol{F}-\boldsymbol{F}^{\dagger}),  \\
\boldsymbol{\Gamma}_{22}&=& -\boldsymbol{F}-\boldsymbol{F}^{\dagger}+\boldsymbol{C}-\boldsymbol{C}^{T}.
\end{eqnarray}
At this stage we have all the ingredients to build the matrix $\boldsymbol{\Gamma}$  and consequently
to calculate the entanglement entropy from Eq.~(\ref{ent2}). In fact, one can calculate the entanglement from
\begin{equation}\label{VTI_mat}
S=-\sum_{j=1}^{l}\left[\frac{1+\nu_{j}}{2}\ln\left(\frac{1+\nu_{j}}{2}\right)+\frac{1-\nu_{j}}{2}\ln\left(\frac{1-\nu_{j}}{2}\right)
\right],
\end{equation} 
where $\nu_{j}$'s are the positive eigenvalues of the matrix $\boldsymbol{\Gamma}$.

Let us now pass to summarize the numerical results illustrated in Fig.~\ref{fig:ent}a and Fig.~\ref{fig:ent}b. 
Firstly, it is  evident that, as long
as $p>1$ and $\gamma\not=0$, $v_g$ does not necessarily  play a role into the time evolution of the entanglement entropy. 
See  Fig.~\ref{fig:ent}a and compare the blue curve, corresponding to $S(t)$ for $p=1$ (i.e. a fully polarized initial state $|\psi_0\rangle=|\uparrow\uparrow\dots\uparrow\rangle$) with the  green that
refers instead to $p=3$; in both cases $\gamma=2$ and $h=1.5$. It is clear that starting from an initial state with
$p\not=1$, the saturation time is state-dependent and larger than $\frac{l}{2v_g}$.

Mimicing the original interpretation in~\cite{CC2006, CC2007}, it  is tempting to conjecture that $\tau_s$ could be approximated instead by
$\frac{l}{2v_{\max}}$, with $v_{\max}$ the light-cone velocity extracted from the correlation functions in Sec.~\ref{sec3}. 
This natural conjecture can be readily verified, for instance,
in the free fermion case  $\gamma=h=0$ and $J=-1$.
In Fig.~\ref{fig:ent}b we report numerical simulations at $\gamma=h=0$ and $J=-1$  for $S(t)$ starting different initial states. 
From top to bottom the configurations that label the selected initial states are $(1/p,1)$,
with $p=2,3$ and $(1/2,2)$. 
In the free fermion case the saturation time is then fully compatible with the conjecture $\tau_s=\frac{l}{2v_{\max}}$ and $v_{\max}$ given in
Eq. \eqref{XX max group velocity} and below Eq.~\eqref{fln_terms2 XX 2}. In particular, the agreement for the
configuration $(\frac{1}{2},2)$ is remarkable because the
result $v_{\max}=\sqrt{2}$ was derived in Sec.~\ref{secff} by using  $A_{22}=0$.

We have done a similar analysis for several different values of $\gamma$ and $h$ to check whether the conjecture for $\tau_s$
holds more generally in the XY chain. The final answer is negative as can be understood again
from Fig.~\ref{fig:ent}a, analyzing in particular the red and green curves. The green curve describes  $S(t)$ for
$p=3$ ($h=1.5$ and $\gamma=2$) where $v_{\max}$ is obtained from Eq.~\eqref{vLC} and shows that
the saturation time is again compatible with
$\frac{l}{2v_{\max}}$. The red curve describes instead $S(t)$ for $p=2$ ($h=1.5$ and $\gamma=2$) and
according to the analysis in Sec.~\ref{sec3}, the light-cone velocity $v_{\max}$ 
is again given by Eq.~\eqref{vLC}.  However the entanglement entropy 
displays a saturation time larger than $\frac{l}{2v_{\max}}$.
At present we do not have  an analytical understanding of this effect.
However,
our numerical simulations indicate that $v_{\max}$ sets a lower bound on the saturation time and actually $\tau_s\geq \frac{l}{2v_{\max}}.$

\section{Conclusions}
\label{conc}
In this paper we have analyzed the influence of the initial state on the maximum speed at which correlations can propagate, according to the Lieb-Robinson bound. 
We investigated the XY chain and global quenches from a  class of initial states that are factorized in the local $z$-component of the spin and have a crystalline 
structure. We demonstrated explicitly that  momentum conservation in the crystal  leads to a state-dependent light-cone velocity $v_{\max}$ that rules how fast correlations
spread. For instance, see Sec.~\ref{sec3n} below Eq.~\eqref{vLC}, pairs of quasi-particles can travel after the quench with momenta that are no more opposite.
This effect slows down the signal propagation.

We have given, and checked numerically, analytical predictions for the light-cone velocities for several values of the parameters $\gamma$, $h$ and $p$;
 concrete examples are given in Eq.~\eqref{XX max group velocity} and Eq.~\eqref{vLC}.
We also discussed  an approximation of the fermionic correlations functions in infinite volume that shows,
in agreement with previous results in\cite{CAiry, B2016}, that the behaviour at the
light-cone edge can be characterized by integer powers of  $t^{-1/3}$. In particular this is the case when
the light-cone velocity is a maximum with vanishing second derivative (and non-zero third derivative) of the effective dispersion.
The degree of universality  of the $t^{-1/3}$-scaling and in particular its dependence on the initial state,
however, have been not  clarified yet~\cite{BD, Prosen, S, DW, BPC}. 

We have then studied numerically the evolution of the entanglement entropy and showed that the choice 
of the initial state affects also the saturation time. When completing this paper, a preprint\cite{BTC} appeared that 
analyzes entanglement dynamics in the XX chain ($\gamma=0$) for the  class of initial states here labeled as $(1/p,1)$. 
In particular an interesting semiclassical interpretation in terms of entangled $p$-plets of quasi-particles is proposed.  
Our calculations in Sec.~\ref{sec3} for the light-cone velocity in the XX chain are in agreement with such a quasi-particle picture.
It will be important to investigate how this can be adapted to determine the light-cone velocity $v_{\max}$ and the linear growth
 of the entanglement entropy also for $\gamma\not=0$. Our analysis suggests that these observables are not easily predictable on the whole parameters space, 
 therefore  a generalized quasi-particle picture will be likely initial state dependent.
Finally, it will be relevant to study the effect of the initial state on Loschmidt echo and finite size effects \cite{MR2017,Piroli2018}.
\vspace*{0.5cm}

\textbf{Acknowledgement}
We are indebted to Pasquale Calabrese for important observations on the draft of this manuscript.
The work of
K.N. is supported by National Science Foundation under
Grant No.  PHY- 1314295. The work of MAR
was supported in part by CNPq. JV and MAR thank  the INCT-IQ initiative for partial support. JV also thanks LPTHE-Paris VI,  SISSA
and the University of Florence for their kind hospitality.\\

\appendix

\section{Light-cone velocities for $p=2$}
\label{app2}
\begin{figure}[ht] 
\includegraphics[width=0.8\columnwidth,angle =0]{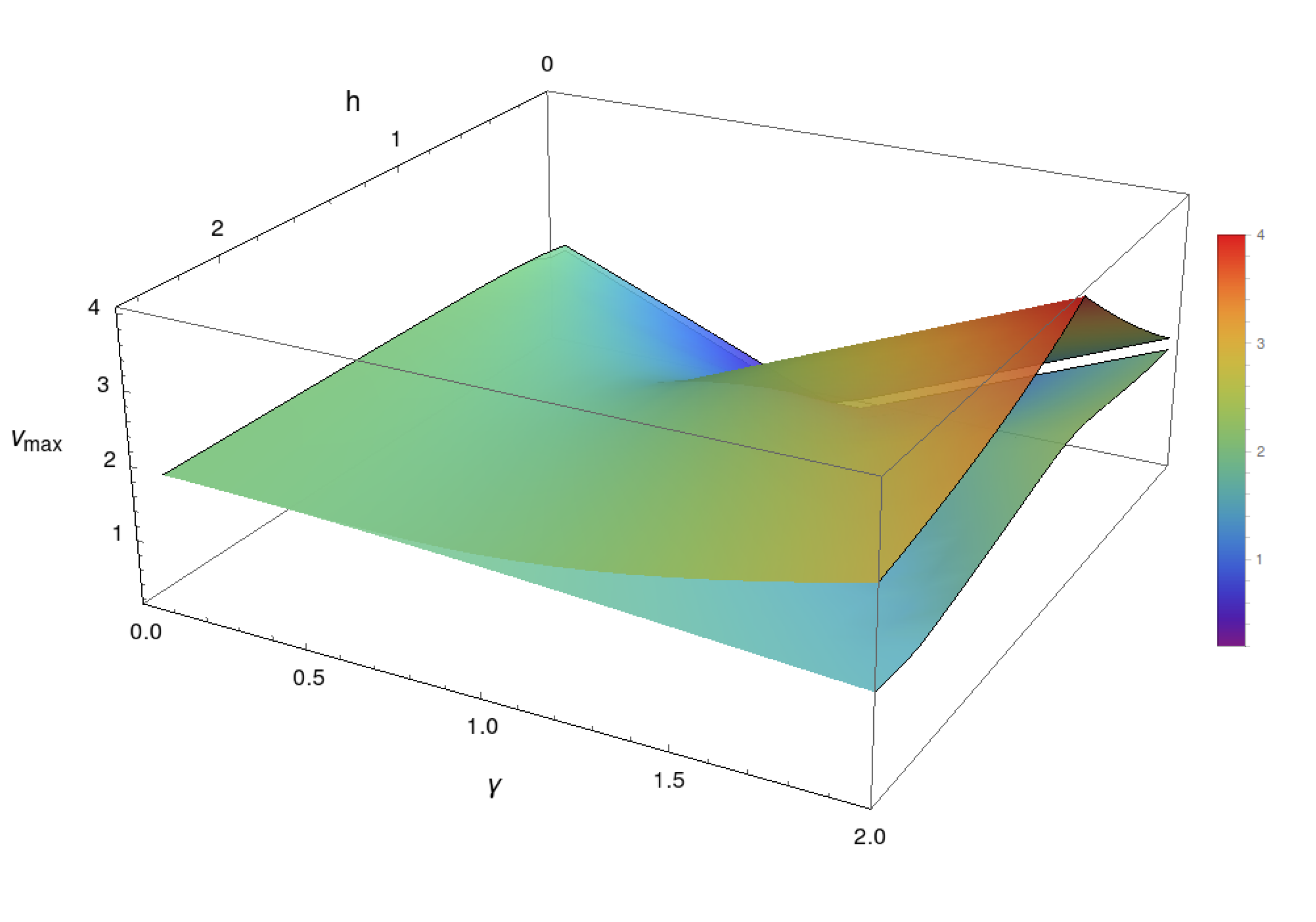}
\includegraphics[width=0.8\columnwidth,angle =0]{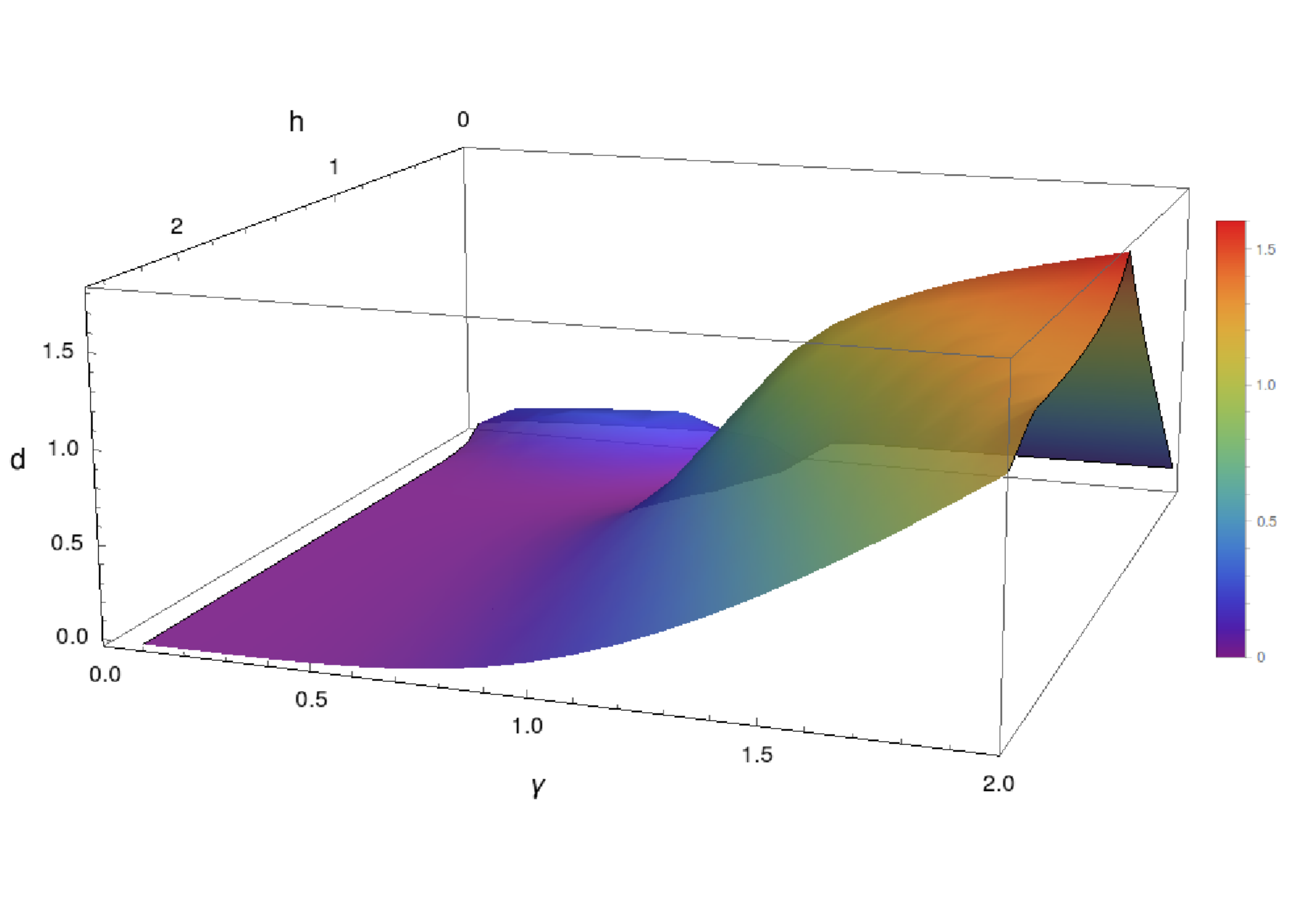}
\caption{(color online) From top to bottom. First panel: Maximum   group velocity $v_{\max}$, obtained from Eq.~\eqref{vLC}  and $v_{g}$ 
for different values of $\gamma$  and $h$ at $p=2$. Second panel: The difference $d$ between $v_{\max}$ and $v_{g}$ at $p=2$. Here we took $J=1$.} 
\label{fig:velocities}
\end{figure}
As we discussed in the main part of the paper
different initial states can induce different effective maximum group velocities. A fully translation invariant initial state with all the spins up or down ($p=1$) leads to a propagation with velocity $v_g$, the maximum of Eq.~\eqref{group velocity}. However, for state $(1/2,1)$ $v_{\max}$ is fixed by Eq.~\eqref{vLC} and  can be easily  found
numerically. In Fig.~\ref{fig:velocities} we plot the light-cone velocities for $p=2$ obtained in this way and compare them  against $v_g$. The figure clearly shows that for small values of $\gamma$
and $h$ the difference between the two are negligible. A difference (denoted as $d$ in Fig.~\ref{fig:velocities}) between the light-cone velocity and $v_g$ is instead more pronounced close to $h=1$. This will be the best condition to test the effects studied in this paper.


\begin{thebibliography}{44}
\bibitem{LRvelocity}The explicit form of the velocity can be written
by first defining the form of the interaction. Consider a quantum spin system defined over a discrete set 
$\Gamma$ with the Hamiltonian $H_{\Lambda}=\sum_{X\subset\Lambda}\Phi(X)$, where $\Lambda\subset\Gamma$. Suppose now a non-increasing function $F:[0,\infty)\to(0,\infty)$ with the following
properties $||F||=\text{Max}_{x\in\Gamma}\sum_{y\in\Gamma}F(|x-y|)<\infty$ and $C=\text{Max}_{x\in\Gamma}\sum_{z\in\Gamma}\frac{F(|x-z|)F(|z-y|)}{F(|x-y|)}<\infty$
and $||\Phi||=\text{Max}_{x,y\in\Gamma}\frac{1}{F(|x-y|)}\sum\limits_{\substack{X\subset\Gamma\\ x,y\in X}}^{}||\Phi(X)||<\infty$. Now for any non-negative real $\mu$
the $F_{\mu}(x)=e^{-\mu x}F(x)$ has the same properties as $F(x)$ and $||F||\geq||F_{\mu}||$ and $C\geq C_{\mu}$. Then the LR velocity can be 
written as $v_{LR}=\text{Max}_{\mu}\frac{2||\Phi_{\mu}||C_{\mu}}{\mu}$.
\bibitem{foot} It is easy to prove that the Wick theorem applies to any eigenstate $|\psi_0\rangle$ of the local $\sigma_l^z$' s operators.
Define $d_{l}$ annihilation operators as $c_{l}$ if $l$ is in the set of coordinates of an up spin and $c^{\dagger}_l$ is vice-versa $l$ is not in such a set.
Analogously define creation operators $d^{\dagger}_l$. It is clear that $|\psi_0\rangle$ is the vacuum of the fermions $d_l$, i.e. $d_l|\psi_0\rangle=0$, $\forall l=1,\dots,L$ and
that the $d_l$'s and $d^{\dagger}_l$'s satisfy canonical anticommutation relations. Therefore the Wick theorem applies to correlation functions of the $d_l$'s and $d^{\dagger}_l$'s
fermions on the state $|\psi_0\rangle$ and then also to correlation functions of the $c_l$'s and $c^{\dagger}_l$'s.
\end{thebibliography}
\end{document}